\documentclass[final,3p,times,sort&compress,doublespacing,fleqn]{elsarticle}

\makeatletter
\def\ps@pprintTitle{%
   \let\@oddhead\@empty
   \let\@evenhead\@empty
   \let\@oddfoot\@empty
   \let\@evenfoot\@oddfoot
}
\makeatother

\usepackage{natbib}

\usepackage{booktabs}

\usepackage{multirow,setspace}

\usepackage{amsthm,amssymb,array,float,latexsym,amsmath,amssymb,amsbsy,eurosym,theorem,amsfonts,mathrsfs,bbm,verbatim}

\usepackage[]{graphicx}
\usepackage{color,soul}

\usepackage{rotating}
\usepackage{tabularx}
\usepackage[table]{xcolor}

\usepackage[many]{tcolorbox}

\usepackage{todonotes}

\usepackage{placeins}

\renewcommand{\b}[1]{\boldsymbol{#1}} 

\renewcommand{\c}[1]{\mathcal{#1}}
\newcommand{\s}[1]{\mathscr{#1}}

\renewcommand{\v}[1]{\text{\usefont{U}{bbm}{m}{n}#1}} 
\newcommand{\oo}[1]{\texttt{\textit{#1}}}

\newcommand{\lrp}[1]{\left( {#1} \right)}
\newcommand{\lb}{\left[}
\newcommand{\rb}{\right]}
\newcommand{\lrb}[1]{\left[ {#1} \right]}

\newcommand{\p}{\partial}

\newcommand{\f}{\displaystyle\frac}

\newcommand{\bnull}{\b0}

\renewcommand{\d}{\mbox{d}}

\newcommand{\Grad}{\ensuremath{\mbox{Grad}}}

\newcommand{\Div}{\mbox{Div}}

\newcommand{\dyad}{\otimes}

\newcommand{\inv}{{}^{\text{\footnotesize -}\text{\scriptsize 1}}}
\newcommand{\trns}{{}^{\text{\scriptsize t}}}

\newcommand{\Det}{\text{Det}}
\newcommand{\Cof}{\text{Cof}}

\newcommand{\sth}{\{\bullet\}}

\newcommand{\n}{{}^{\text{\tiny{/}}}}

\newcommand{\xt}{{}^{^{\mbox{\scriptsize ext}}}}
\newcommand{\nt}{{}^{^{\mbox{\scriptsize int}}}}

\renewcommand{\i}{ {}^{^{|}} }
\newcommand{\ii}{ {}^{^{||}} }
\newcommand{\iii}{ {}^{^{|||}} }

\newcommand{\pp}{ {}_{_{\text{\scriptsize{1}}}} }
\newcommand{\ppp}{ {}_{_{\text{\scriptsize{2}}}}  }
\newcommand{\pppp}{ {}_{_{\text{\scriptsize{3}}}} }

\newcommand{\ppi}{ \pp \!\! \i }
\newcommand{\pppi}{ \ppp \!\! \i }

\newcommand{\ppppi}{ \pppp \!\! \i }

\newcommand{\pppinii}{ \ppp \!\! \i\n\ii }
\newcommand{\ppppiniiniii}{ \pppp \!\! \i\n\ii\n\iii }

\newdefinition{rmk}{Remark}

\begin{document}

\begin{frontmatter}

\title{The computational framework for continuum-kinematics-inspired peridynamics}

\author[bilkent]{A.~Javili\corref{cor}}
\ead{ajavili@bilkent.edu.tr}
\author[bilkent]{S.~Firooz}
\author[glasgow]{A.~T.~McBride}
\author[glasgow,erlangen]{P.~Steinmann}
\address[bilkent]{Department of Mechanical Engineering, Bilkent University, 06800 Ankara, Turkey}
\address[glasgow]{Glasgow Computational Engineering Centre, James Watt School of Engineering, University of Glasgow, Glasgow G12 8QQ, United Kingdom}
\address[erlangen]{Chair of Applied Mechanics, University of Erlangen-Nuremberg, Egerland Str. 5, 91058 Erlangen, Germany}
\cortext[cor]{Corresponding author.}

\begin{abstract}
Peridynamics (PD) is a non-local continuum formulation.
The original version of PD was restricted to bond-based interactions.
Bond-based PD is geometrically exact and its kinematics are similar to classical continuum mechanics (CCM).
However, it cannot capture the Poisson effect correctly.
This shortcoming was addressed via state-based PD, but the kinematics are not accurately preserved.
Continuum-kinematics-inspired peridynamics (CPD) provides a geometrically exact framework whose underlying kinematics coincide with that of CCM and captures the Poisson effect correctly.
In CPD, one distinguishes between one-, two- and three-neighbour interactions.
One-neighbour interactions are equivalent to the bond-based interactions of the original PD formalism.
However, two- and three-neighbour interactions are fundamentally different from state-based interactions as the basic elements of continuum kinematics are preserved precisely.
The objective of this contribution is to elaborate on computational aspects of CPD and present detailed derivations that are essential for its implementation.
Key features of the resulting computational CPD are elucidated via a series of numerical examples.
These include three-dimensional problems at large deformations.
The proposed strategy is robust and the quadratic rate of convergence associated with the Newton--Raphson scheme is observed.
\end{abstract}


\end{frontmatter}

\section{Introduction}\label{sec:intro}

Peridynamics is an alternative approach to formulate non-local continuum mechanics~\cite{0-Silling2000}; its roots can be traced back to the pioneering works of Piola~\cite{DellIsola2015,DellIsola2017}.
However, it is fundamentally different from established non-local elasticity~\citep[see][among others]{Eringen2002,Forest2009} as the concepts of stress and strain are absent.
As a non-local theory, the behaviour of each material point is influenced by interactions with other material points in their finite vicinity.
In contrast to CCM, the governing equations of PD are integro-differential equations appropriate for problems involving discontinuities such as cracks and interfaces.
Given that PD inherently accounts for geometrical discontinuities, it provides a suitable framework for fracture mechanics and related problems~\cite{53-Kilic2009a,145-Foster2011,76-Silling2010,126-Agwai2011,264-Dipasquale2014,339-Chen2015a,500-Han2016,470-Emmrich2016,512-DeMeo2016,464-Sun2016,478-Diyaroglu2016,Giannakeas2020}.
However, the range of PD applications is broad and not limited to fracture.
PD has experienced prolific growth as an area of research, with a significant number of contributions in multiple disciplines.
Various applications and extensions of PD not dealing exclusively with material failure include quasi-static problems~\cite{11-Dayal2006,178-Mikata2012,314-Breitenfeld2014,345-Huang2015,448-Madenci2016a}, coupled problems~\cite{44-Gerstle2008,89-Bobaru2010,Oterkus2014a,Oterkus2014,580-Oterkus2017}, multiscale modeling~\cite{50-Bobaru2009,114-Shelke2011,265-Rahman2014,Talebi2014,358-Ebrahimi2015,510-Tong2016,491-Xu2016,Zaccariotto2018}, structural mechanics~\cite{5-Silling2005,288-OGrady2014a,364-Taylor2015,472-Chowdhury2016,Li2016}, constitutive models~\cite{285-Aguiar2014,328-Sun2014,Silhavy2017,576-Madenci2017,Silling2017b,Chen2016b,Wang2019,Pathrikar2019}, biomechanics~\cite{523-Taylor2016,Lejeune2017a}, and wave dispersion~\cite{123-Zingales2011,Vogler2012,268-Wildman2014,522-Bazant2016a,449-Nishawala2016,526-Silling2016,Butt2017}.
For an extensive study of the balance laws, applications, and implementations, see \cite{Madenci2014}.
For a brief description of PD together with a review of its applications and related studies in different fields to date, see~\cite{Javili2018a}.
Very recently, Bode et al.~\cite{Bode2020,Bode2020a} proposed a mixed PD formulation as a generalization of PD theory that offers a stable alternative suitable for finite deformations, also referred to as Peridynamic Petrov--Galerkin method.
Fundamental works on PD are growing in number but are still relatively limited, see e.g.\ \cite{92-Silling2010b,Ostoja-Starzewski2013}.
Note, that while the discretised format of PD bears a similarity to discrete mechanics, it is still a continuum formulation.
For further connections and differences between PD theory, continuum mechanics and particle systems, see the  contributions~\cite{Fried2010,47-Murdoch2012,Fosdick2013,Podio-Guidugli2017}, among others.

The original PD theory of Silling~\cite{0-Silling2000} was restricted to bond-based interactions.
This limited its applicability for material modelling.
For example, the theory was unable to account for a Poisson ratio other than $1/4$ for isotropic materials.
This shortcoming was addressed in various contributions and finally rectified in \cite{16-Silling2007} via the introduction of the notion of \emph{state} and the categorising of interactions as bond-based, ordinary state-based or non-ordinary state-based.
The kinematics in state-based PD do not exactly follow the motion of a continuum body and can lead to non-physical deformation modes or instabilities, as discussed in~\citep{313-Tupek2014,Silling2017a} in the context of \emph{correspondence}.
Recently, Javili~et~al.~\cite{Javili2019} introduced a \emph{continuum-kinematics-inspired} approach to peridynamics, referred to as CPD, which bridges the gap between CCM and PD.
More precisely, CPD is an alternative PD formulation whose underlying kinematics exactly follows the motion of a continuum body and in the limit coincides with those of CCM.
The interaction potentials in CPD are decomposed into three parts corresponding to \emph{one-}, \emph{two-} and \emph{three-neighbour} interactions within a horizon.
One-neighbour interactions are equivalent to the bond-based interactions of the original PD formalism.
However two- and three-neighbour interactions are fundamentally different from state-based interactions.
\textit{The main objective of this manuscript is to elaborate on the computational aspects of CPD and to provide detailed numerical examples at finite deformations to illustrate the theory and demonstrate its potential.}

The manuscript is organised as follows.
Section~\ref{sec:definition} introduces the notation, elaborates on the kinematics of the problem, presents the governing equations of CPD and provides suitable constitutive laws that inherently account for material frame indifference via dependence on objective deformation measures.
Next, we detail the implicit computational implementation of the governing equations and their approximate forms in Section~\ref{sec:implementation}.
In particular, we discuss the computational aspects for treating a finite deformation, quasi-static problem.
To do so, we propose physically meaningful interaction energy densities and provide detailed derivations.
Thereafter, in Section~\ref{sec:examples}, the key features of the interaction energies and capabilities of CPD are illustrated via a series of numerical examples.
Section~\ref{sec:conclusion} concludes this work and provides an outlook.

\section{Problem definition}\label{sec:definition}

This section briefly defines the problem of CPD and gathers its main relations and equations.
Central to CPD is the kinematic description as detailed in Section~\ref{sec:kinematics}.
This is inspired by CCM.
Thereafter, the key quantities of CPD together with the governing equations are given in Section~\ref{sec:governing}.
Thermodynamically consistent constitutive laws for CPD are derived in Section~\ref{sec:constitutive}.
In Section~\ref{sec:angmom}, we propose specific forms of constitutive laws for one-neighbour, two-neighbour and three-neighbour interactions in a unified format.

\subsection{Kinematics}\label{sec:kinematics}

\begin{figure}[h]
    \centering
	\includegraphics[width=1.0\textwidth]{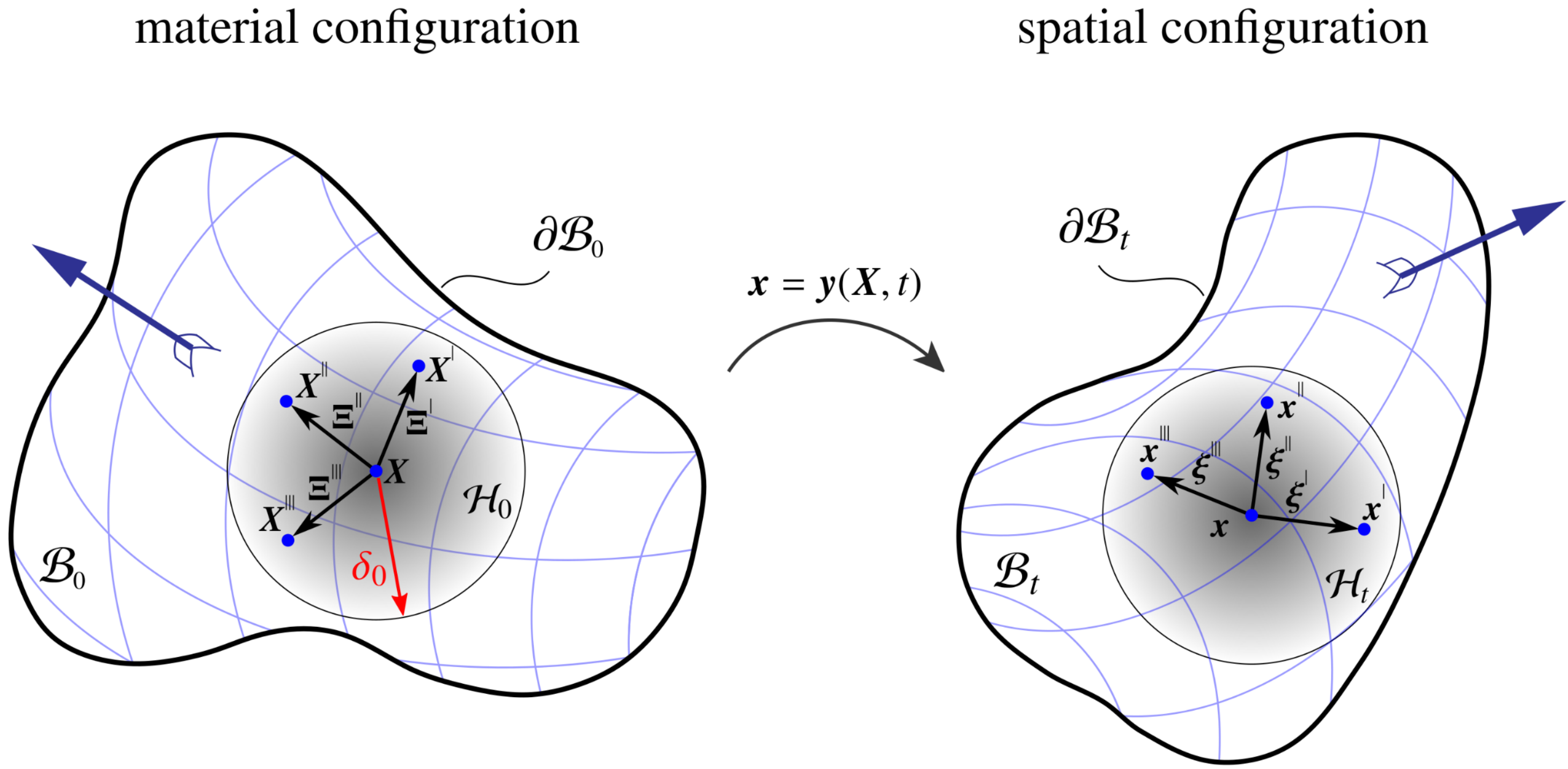}
	\caption{Deformation of a continuum body within the CPD formulation.
	The continuum body that occupies the material configuration $\c{B}_0 \subset \v{R}^3$ at time $t=0$ is mapped to the spatial configuration $\c{B}_t \subset \v{R}^3$ via the nonlinear deformation map $\b{y}$.
	The neighbourhood of $\b{X}$ is mapped onto the neighbourhood of $\b{x}$.
	That is, the neighbour set $\{\b{X}\i, \b{X}\ii, \b{X}\iii\}$ is mapped onto $\{\b{x}\i, \b{x}\ii, \b{x}\iii\}$, respectively.
	}
    \label{fig:motion}
\end{figure}

Consider a continuum body that occupies the material configuration $\c{B}_0 \subset \v{R}^3$ at time $t=0$ and is mapped to the spatial configuration $\c{B}_t \subset \v{R}^3$ via the nonlinear deformation map $\b{y}$ as $\b{x} = \b{y} (\b{X},t) : \c{B}_0 \times \v{R}_+ \to \c{B}_t$
with $\b{X}$ and $\b{x}$ identifying the points in the material and spatial configurations, respectively, as illustrated in Fig.~\ref{fig:motion}.
Note that we restrict the analysis to quasi-static conditions.
Thus time plays the role of a history parameter to order the sequence of events.
Central to the CPD theory, and in contrast to standard local continuum mechanics, is the non-locality assumption that any point $\b{X} \in \c{B}_0$ can interact with points within its finite neighbourhood $\c{H}_0(\b{X})$.
The neighbourhood $\c{H}_0$ is referred to as the \emph{horizon} in the material configuration.
The \emph{measure} of the horizon in the material configuration is denoted by $\delta_0 := \text{meas}(\c{H}_0)$ and is generally the radius of a spherical neighbourhood centred at $\b{X}$.
The spatial horizon $\c{H}_t= \b{y} (\c{H}_0(\b{X}),t)$ is the image of the material horizon $\c{H}_0$ under the deformation map $\b{y}$ and in general will not remain spherical.
Note that the horizons $\c{H}_0$ and $\c{H}_t$ coincide with the points $\b{X}$ and $\b{x}$ in the limit of an infinitesimal neighbourhood, thereby recovering the kinematics of the local continuum mechanics formalism.

To be more precise, we identify the neighbours within the horizon by a superscript.
For instance, the point $\b{X}\i \in \c{H}_0(\b{X})$ denotes a neighbour of point $\b{X}$ in the material configuration.
The point $\b{x}\i$ within the horizon of $\b{x}$ is the spatial counterpart of the point $\b{X}\i$ defined through the nonlinear deformation map $\b{y}$ as $\b{x} \i := \b{y} ( \b{X}\i , t )$.
In our proposed framework, for any point $\b{X}$ we identify all  possible neighbour sets $\{\b{X}\i, \b{X}\ii, \b{X}\iii\}$ that are mapped onto  $\{\b{x}\i, \b{x}\ii, \b{x}\iii\}$, respectively, as shown in Fig.~\ref{fig:motion}.
The relative positions, i.e.\ the finite line elements, in the material and spatial configurations are denoted as $\b{\Xi}^{\sth}$ and $\b{\xi}^{\sth}$, respectively, where the superscript $\sth$ identifies the neighbour; that is
\begin{equation}
\begin{aligned}
	& \b{\Xi} \i := \b{X}\i - \b{X} && \qquad \text{and} \qquad && \b{\xi} \i := \b{x}\i - \b{x} && \qquad \text{where} \qquad &&  \b{\xi}\i = \b{\xi} ( \b{X}\i ; \b{X} ) = \b{y}(\b{X}\i) - \b{y}(\b{X}) \,, \\
	& \b{\Xi} \ii := \b{X}\ii - \b{X} && \qquad \text{and} \qquad && \b{\xi} \ii := \b{x}\ii - \b{x} && \qquad \text{where} \qquad &&  \b{\xi}\ii = \b{\xi} ( \b{X}\ii ; \b{X} ) = \b{y}(\b{X}\ii) - \b{y}(\b{X}) \,, \\
	& \b{\Xi} \iii := \b{X}\iii - \b{X} && \qquad \text{and} \qquad && \b{\xi} \iii:= \b{x}\iii - \b{x} && \qquad \text{where} \qquad &&  \b{\xi}\iii = \b{\xi} ( \b{X}\iii ; \b{X} ) = \b{y}(\b{X}\iii) - \b{y}(\b{X}) \,.
\end{aligned}
\end{equation}

Before defining the kinematic measures of CPD, we recall the three \emph{local kinematic measures} of relative deformation in CCM, namely the deformation gradient $\b{F}:= \Grad \, \b{y}$, its cofactor $\b{K}:= \Cof \, \b{F}$ and its determinant $J:= \Det \, \b{F}$.
In the spirit of these local measures, we introduce three \emph{non-local kinematic measures} of relative deformation, namely $\b{\xi} \i$, $\b{a}\i\n\ii$ and $v\i\n\ii\n\iii$ associated with CPD.
The \emph{first relative deformation measure} of CPD is $\b{\xi} \i$.
 It mimics the linear map $\b{F}$ from the infinitesimal line element $\d \b{X}$ in the material configuration to its spatial counterpart $\d \b{x}$.
In view of our proposed CPD formalism, the relative deformation measure
\begin{equation}
\begin{aligned}
 	\b{\xi} \i=\b{x}\i - \b{x} \,,
\end{aligned}
\end{equation}
is the main ingredient to describe \emph{one-neighbour interactions}.
The \emph{second relative deformation measure} $\b{a}\i\n\ii$ is reminiscent of the linear map $\b{K}$ from the infinitesimal vectorial area element $\d\b{A}$ in the material configuration to its spatial counterpart $\d\b{a}$.
This is essentially Nanson's formula from conventional continuum kinematics.
In our proposed framework, the relative area measure
\begin{equation}
\begin{aligned}
 	\b{a}\i\n\ii=[ \b{x}\i - \b{x} ] \times [ \b{x}\ii - \b{x} ] \,,
\end{aligned}
\end{equation}
is the main ingredient to describe \emph{two-neighbour interactions}.
The \emph{third relative deformation measure} ${v}\i\n\ii\n\iii$ mimics the linear map $J$ from the infinitesimal volume element $\d {V}$ in the material configuration to its spatial counterpart $\d {v}$.
The relative volume measure
\begin{equation}
\begin{aligned}
	{v}\i\n\ii\n\iii = \Big[ [ \b{x}\i - \b{x} ] \times [ \b{x}\ii - \b{x} ] \Big]\cdot [ \b{x}\iii - \b{x} ]  \,,
\end{aligned}
\end{equation}
is the main ingredient to describe \emph{three-neighbour interactions}.

\subsection{Governing equations}\label{sec:governing}

Equipped with the kinematic description of CPD, we  briefly recall the governing equations.
Similar to CCM, the governing equations of CPD can be expressed in global or point-wise form.
In contrast to CCM, the point-wise equations of CPD are not local.
That is, applying a localization procedure on global forms in CPD renders point-wise relations that still contain integrals over the horizon and are thus non-local.
It is sometimes possible to apply a localization procedure on the non-local forms of CPD that yield neighbour-wise equations valid for each pair of neighbouring points that are not integrals and hence local.
The global form of the linear momentum balance for quasi-static problems reads
\begin{equation}\label{eq:gov-1}
\begin{aligned}
	\int_{\p\c{B}_0} \b{t}\xt_0 \, \d A + \int_{\c{B}_0} \b{b}\xt_0 \, \d V = \bnull \,,
\end{aligned}
\end{equation}
where $\b{b}\xt_0$ denotes the external force density per volume in the material configuration, with units $\mathrm{N}/\mathrm{m}^3$, and $\b{t}\xt_0$ is the external traction on the boundary in the material configuration, with units $\mathrm{N}/\mathrm{m}^2$.
This format of the external loading is a particular sub-case of a more general case accounting for higher-gradient and non-local continua as detailed in~\citep{Javili2013c,Auffray2015} among others.
The universal form of the quasi-static linear momentum balance~(\ref{eq:gov-1})  can alternatively be expressed in terms of volume integrals as
\begin{equation}\label{eq:gov-2}
\begin{aligned}
	\int_{\c{B}_0} \b{b}\nt_0 \, \d V  + \int_{\c{B}_0} \b{b}\xt_0 \, \d V = \bnull \,.
\end{aligned}
\end{equation}
The internal body force density in the material configuration $\b{b}\nt_0$ in CCM is the material \emph{divergence} of the Piola stress $\b{P}$.
In CPD however, $\b{b}\nt_0$ takes an \emph{integral} form over the horizon.
That is
\begin{equation}\label{eq:gov-3}
\begin{aligned}
	\b{b}\nt_0 = \int_{\c{H}_0} \b{p}\i \, \d V\i \,,
\end{aligned}
\end{equation}
with $\b{p}\i$  the force density per volume squared, with units $\mathrm{N}/\mathrm{m}^6$.
Inserting the internal body force density~(\ref{eq:gov-3}) into the quasi-static linear momentum balance~(\ref{eq:gov-2}) yields, after localization, the \emph{linear momentum balance of CPD}
\begin{equation}\label{eq:TD_BOLM_LOC}
\begin{aligned}
	\b{b}\nt_0 + \b{b}\xt_0 = \bnull \qquad \Rightarrow \qquad \int_{\c{H}_0} \b{p}\i \, \d V\i + \b{b}\xt_0 = \bnull \,.
\end{aligned}
\end{equation}
To derive the angular momentum balance, we start from the global form of the quasi-static moment balance
\begin{equation}
\begin{aligned}
	\int_{\p\c{B}_0} \b{y} \times \b{t}\xt_0 \, \d A + \int_{\c{B}_0} \b{y} \times \b{b}\xt_0 \, \d V = \bnull \,,
\end{aligned}
\end{equation}
which, after some mathematical steps and using the linear momentum balance~(\ref{eq:gov-1}), upon localization reduces to  the \emph{angular momentum balance of CPD}
\begin{equation}\label{eq:gov-5}
\begin{aligned}
	\int_{\c{H}_0} \b{\xi}\i \times \b{p}\i \, \d V\i = \bnull  \,.
\end{aligned}
\end{equation}
Table~\ref{tab:all-compare} gathers the key governing equations for both CCM and CPD for the case of quasi-statics.

\begin{table}[htb]
\caption{Governing quasi-static equations of classical continuum mechanics (CCM) and continuum-kinematics-inspired peridynamics (CPD).
The Piola stress is denoted by $\b{P}$ and $\b{F} \cdot \b{P}\trns$ is symmetric due to angular momentum balance expressed using the third-order permutation tensor $\b{\varepsilon}$.
}
\setstretch{2.4}
\begin{tabular}{|>{\centering\arraybackslash}p{0.22\textwidth}|>{\centering\arraybackslash}p{0.35\textwidth}|>{\centering\arraybackslash}p{0.36\textwidth}|}
    \hline
     & linear momentum balance & angular momentum balance \\
    \hline \hline
    CCM & $\displaystyle \Div \b{P} + \b{b}\xt_0 = \bnull$ &  $\displaystyle \b{\varepsilon} : [ \b{F} \cdot \b{P}\trns ] = \bnull$ \\[6pt]
    \hline
    CPD & $\displaystyle \int_{\c{H}_0} \b{p}\i \, \d V\i + \b{b}\xt_0= \bnull$ &  $\displaystyle \int_{\c{H}_0} \b{\xi}\i \times \b{p}\i \, \d V\i = \bnull$ \\[6pt]
    \hline
\end{tabular}
\label{tab:all-compare}
\end{table}

\subsection{Constitutive laws}\label{sec:constitutive}

Constitutive laws bridge the gap between the kinematics described in Section~\ref{sec:kinematics} and the kinetics in Section~\ref{sec:governing}.
The constitutive laws of CPD, as in CCM, must be thermodynamically consistent and are thus derived via a Coleman--Noll-like procedure.
Let $\Psi$ denote the point-wise stored energy density per volume in the material configuration.
The dissipation power density $\s{D}$ reads
\begin{equation}\label{eq:constitutive-1}
\begin{aligned}
	\s{D} = \int_{\c{H}_0} \b{p}\i \cdot \dot{\b{\xi}}\i \, \d V\i - \dot{\Psi} \geq 0 \,.
\end{aligned}
\end{equation}
For elasticity $\s{D}=0$, and the inequality~(\ref{eq:constitutive-1}) becomes
\begin{equation}\label{eq:constitutive-2-}
\begin{aligned}
	\dot{\Psi} = \int_{\c{H}_0} \b{p}\i \cdot \dot{\b{\xi}}\i  \, \d V\i \,.
\end{aligned}
\end{equation}
This relation forms the basis for the derivation of the hyperelastic constitutive laws in CPD.
The stored energy density $\Psi$ consists of the contributions from one-, two- and three-neighbour interactions.
Let $\psi\ppi$, $\psi\pppinii$ and $\psi\ppppiniiniii$ denote the stored energy densities corresponding to one-neighbour, two-neighbour and three-neighbour interactions, respectively, in the material configuration.
That is
\begin{equation}\label{eq:intr-pot}
\begin{aligned}
	& \psi\ppi = \psi\pp(\b{\xi}\i)  && : && \text{one-neighbour interaction energy density} && , && \lrb{ \psi\ppi } =\textrm{N}.\textrm{m}/\textrm{m}^6 \,, \\
	& \psi\pppinii = \psi\ppp(\b{a}\i\n\ii) && : && \text{two-neighbour interaction energy density} && , && \lrb{ \psi\pppinii } =\textrm{N}.\textrm{m}/\textrm{m}^9 \,, \\
	& \psi\ppppiniiniii = \psi\pppp(v\i\n\ii\n\iii) && : && \text{three-neighbour interaction energy density} && , && \lrb{ \psi\ppppiniiniii } =\textrm{N}.\textrm{m}/\textrm{m}^{12} \,.
\end{aligned}
\end{equation}
The point-wise stored energy density $\Psi$ reads
\begin{equation}\label{eq:varPsi_dens}
\begin{aligned}
	& \Psi = \int_{\c{H}_0}  {\f{1}{2} \, \psi\ppi} \d V\i + \int_{\c{H}_0} { \int_{\c{H}_0} \f{1}{3} \, \psi\pppinii \, \d V\ii \d V\i} + \int_{\c{H}_0}  {\int_{\c{H}_0} \int_{\c{H}_0} \f{1}{4} \, \psi\ppppiniiniii \, \d V\iii \d V\ii \d V\i} \,,
\end{aligned}
\end{equation}
where the factors one-half, one-third and one-fourth are required to prevent multiple counting of energy since we visit each point multiple times depending on the number of integrals.
Thus, the rate of the stored energy density $\dot\Psi$ reads
\begin{equation}\label{eq:varPsi_dens-1}
\begin{aligned}
	\dot{\Psi} = \int_{\c{H}_0} \lrb{ \b{p}\i_1 + \b{p}\i_2 + \b{p}\i_3 } \cdot \dot{\b{\xi}}\i \; \d V\i \,,
\end{aligned}
\end{equation}
in which the vectors $\b{p}\i_1$, $\b{p}\i_2$ and $\b{p}\i_3$ are defined by
\begin{equation}\label{eq:varPsi_dens-2}
\begin{aligned}
	& \b{p}\i_1 := {\f{\p \psi\ppi}{\p \b{\xi}\i}} \qquad , \qquad \b{p}\i_2 := {\int_{\c{H}_0} 2 \, \b{\xi}\ii \times \f{\p \psi\pppinii}{\p \b{a}\i\n\ii} \, \d V\ii} \qquad , \qquad \b{p}\i_3 :=  {\int_{\c{H}_0} \int_{\c{H}_0}  3 \, \b{\xi}\ii \times \b{\xi}\iii  \, \f{\p \psi\ppppiniiniii}{\p v\i\n\ii\n\iii} \; \d V\iii \d V\ii} \,,
\end{aligned}
\end{equation}
the derivation of which is omitted here for the sake of brevity \citep[see][for further details]{Javili2019}.
It is important to keep in mind that to arrive at the definition~(\ref{eq:varPsi_dens-2})$_2$, we require ${\p \psi\i}/{\p \b{a}\i\n\ii}$ to be homogeneous of degree one in ${\b{a}\i\n\ii}$.
Note that at this stage, $\b{p}\i_1$, $\b{p}\i_2$ and $\b{p}\i_3$ in Eq.~(\ref{eq:varPsi_dens-2})  simply provide a structure for a constitutive relation and do not necessarily imply a  particular physical meaning.
Inserting the rate of the stored energy density~(\ref{eq:varPsi_dens-1}) into equality~(\ref{eq:constitutive-2-}), immediately reveals the \emph{hyperelastic constitutive law of CPD} as
\begin{equation}\label{eq:force-densities}
\begin{aligned}
	\b{p}\i = \b{p}\i_1 + \b{p}\i_2 + \b{p}\i_3 \,.
\end{aligned}
\end{equation}
This clearly implies that $\b{p}\i_1$, $\b{p}\i_2$ and $\b{p}\i_3$ can be interpreted as force density vectors due to one-neighbour, two-neighbour and three-neighbour interactions, respectively and hence the notation employed.

\subsection{Examples of constitutive laws}\label{sec:angmom}

In this section, we provide both generic and specific examples of hyperelastic constitutive laws for CPD.
That is, we give a concrete form for the interaction potentials~(\ref{eq:intr-pot}).
The specific examples have not been presented to date in the literature.
We investigate the possible options and seek interaction potentials that \textit{a priori} fulfil the CPD angular momentum balance~(\ref{eq:gov-5}).
That is
\begin{equation}\label{eq:angmom-2}
\begin{aligned}
	\int_{\c{H}_0} \b{\xi}\i \times \b{p}\i \, \d V\i \stackrel{!}{=} \bnull  \qquad \Rightarrow \qquad \int_{\c{H}_0} \b{\xi}\i \times \b{p}\ppi \ \d V\i + \int_{\c{H}_0} \b{\xi}\i \times \b{p}\pppi \ \d V\i + \int_{\c{H}_0} \b{\xi}\i \times \b{p}\ppppi \ \d V\i \stackrel{!}{=} \bnull  \,.
\end{aligned}
\end{equation}
The angular momentum balance~(\ref{eq:angmom-2}) is \emph{sufficiently} satisfied if each of the three integrals vanish identically.
That is
\begin{equation}\label{eq:angmom-3}
\begin{aligned}
	\int_{\c{H}_0} \b{\xi}\i \times \b{p}\ppi \ \d V\i \stackrel{!}{=} \bnull \qquad , \qquad \int_{\c{H}_0} \b{\xi}\i \times \b{p}\pppi \ \d V\i \stackrel{!}{=} \bnull \qquad , \qquad \int_{\c{H}_0} \b{\xi}\i \times \b{p}\ppppi \ \d V\i \stackrel{!}{=} \bnull \,.
\end{aligned}
\end{equation}
To proceed, we define scalar-valued line, area and volume measures $\oo{l}$, $\oo{a}$ and $\oo{v}$, respectively as
\begin{equation}\label{eq:}
\begin{aligned}
	\oo{l}:= |\b{\xi}\i| \quad , \quad \oo{a}:= |\b{a}\i\n\ii| = |\b{\xi}\i\times\b{\xi}\ii|  \quad , \quad \oo{v}:= |{v}\i\n\ii\n\iii| = |\b{\xi}\i\cdot[\b{\xi}\ii\times\b{\xi}\iii]| \,.
\end{aligned}
\end{equation}
It has been proven~\citep{Javili2019} that if the energy densities~(\ref{eq:intr-pot}) are expressed in terms of $\oo{l}$, $\oo{a}$ and $\oo{v}$ instead of $\b{\xi}\i$, $\b{a}\i\n\ii$ and $v\i\n\ii\n\iii$, respectively, then conditions~(\ref{eq:angmom-3}) are fulfilled \emph{a priori}.
That is, we require
\begin{equation}\label{eq:intr-pot-2}
\begin{aligned}
	& \psi\ppi = \psi\pp(\b{\xi}\i) = \psi\pp(\oo{l}) && , && \psi\pppinii = \psi\ppp(\b{a}\i\n\ii) = \psi\ppp(\oo{a}) && , && \psi\ppppiniiniii = \psi\pppp(v\i\n\ii\n\iii) = \psi\pppp(\oo{v}) \,.
\end{aligned}
\end{equation}
The interaction potentials~(\ref{eq:intr-pot-2}) are generic examples of suitable one-, two- and three-neighbour interactions.

Finally, we propose specific examples of stored energy densities $\psi\i_{1,2,3}$ that are both thermodynamically consistent and satisfy the angular momentum balance.
These energy densities are employed in the numerical examples in Section~\ref{sec:examples} and their key characteristics are illustrated and discussed.
For that it proves convenient to define the counterparts of $\oo{l}$, $\oo{a}$ and $\oo{v}$ in the material configuration, denoted by $\oo{L}$, $\oo{A}$ and $\oo{V}$, respectively.
That is
\begin{equation}\label{eq:angmom-5}
\begin{aligned}
	\oo{L}:= |\b{\Xi}\i| \quad , \quad \oo{A}:= |\b{A}\i\n\ii| = |\b{\Xi}\i\times\b{\Xi}\ii|  \quad , \quad \oo{V}:= |{V}\i\n\ii\n\iii| = |\b{\Xi}\i\cdot[\b{\Xi}\ii\times\b{\Xi}\iii]|  \,.
\end{aligned}
\end{equation}
An example of the stored energy density per volume squared in the material configuration for one-neighbour interactions $\psi\i_1=\psi_1(\oo{l};\oo{L})$ and in accordance with the original bond-based model of Silling~\cite{0-Silling2000} reads
\begin{equation}\label{eq:angmom-6}
\begin{aligned}
	\psi\ppi = \psi_1(\oo{l};\oo{L}) = \frac{1}{2} \, C_1 \, \oo{L} \, \lb S_1 - 1 \rb^2 \qquad \text{with} \qquad \lb C_1 \rb = \f{\mathrm{N}.\mathrm{m}}{\mathrm{m}^7} \qquad \text{and} \qquad S_1 :=  \f{\oo{l}}{\oo{L}} \,,
\end{aligned}
\end{equation}
where $C_1$ is the one-neighbour elastic coefficient and can be viewed as the resistance against the change of length between a point and its neighbours, reminiscent of the elastic modulus in CCM.
Note that the parameter $S_1$ is precisely the stretch of the bond from a point to its first neighbour.
Motivated by the format of the stored energy density for one-neighbour interactions~(\ref{eq:angmom-6}), we propose the stored energy density per volume cubed for two-neighbour interactions as
\begin{equation}\label{eq:angmom-7}
\begin{aligned}
	\psi\pppinii = \psi_2(\oo{a};\oo{A}) = \frac{1}{2} \, C_2 \, \oo{A} \, \lb S_2 - 1 \rb^2  \qquad \text{with} \qquad \lb C_2 \rb = \f{\mathrm{N}.\mathrm{m}}{\mathrm{m}^{11}}  \qquad \text{and} \qquad S_2 :=  \f{\oo{a}}{\oo{A}} \,,
\end{aligned}
\end{equation}
with $C_2$ the two-neighbour elastic coefficient which can be interpreted as the resistance against the change of the area of the triangle formed by a point and a pair of its neighbours, analogous to Poisson-like effects in CCM for two-dimensional manifolds.
The parameter $S_2$ is essentially the area stretch of this triangle.
Similar to the stored energy densities for one- and two-neighbour interactions, we propose the stored energy density per volume to the fourth power for three-neighbour interactions as
\begin{equation}\label{eq:angmom-8}
\begin{aligned}
	\psi\ppppiniiniii = \psi_3(\oo{v};\oo{V}) = \frac{1}{2} \, C_3 \, \oo{V} \, \lb S_3 - 1 \rb^2  \qquad \text{with} \qquad \lb C_3 \rb = \f{\mathrm{N}.\mathrm{m}}{\mathrm{m}^{15}}  \qquad \text{and} \qquad S_3 :=  \f{\oo{v}}{\oo{V}} \,,
\end{aligned}
\end{equation}
where $C_3$ is the three-neighbour elastic coefficient, which can be interpreted as the resistance against the change of the volume of the tetrahedron formed by each point and its triplet of neighbours, analogous to volumetric Poisson-like effects of CCM.
The stored energy densities~(\ref{eq:angmom-6})--(\ref{eq:angmom-8}) are introduced in this fashion since the energy density~(\ref{eq:angmom-6}) is  identical to the common format used in bond-based PD~\cite{7-Silling2005a}.

Alternative formats for energy densities could be proposed and their consequences investigated.
However, the main objective of this manuscript is to provide details of the computational implementation which remains largely independent of the specific format of the stored energy density.
Table~\ref{tab:BOAM-conseq} summarises the discussion of constitutive laws and collects the fundamental relations and definitions of CPD together with generic and specific examples of the stored energy densities that \emph{sufficiently} satisfy the angular momentum balance.
\begin{table}[htb]
\caption{Unification of concepts and fundamental relations of CPD.}
\setstretch{2.4}
\begin{tabular}{|>{\centering\arraybackslash}p{0.22\textwidth}|>{\centering\arraybackslash}p{0.35\textwidth}|>{\centering\arraybackslash}p{0.36\textwidth}|}
    \hline
    one-neighbour& two-neighbour& three-neighbour\\
    \hline \hline
    \multicolumn{3}{|c|}{force density per volume squared in the material configuration with dimension $\mathrm{N}/\mathrm{m}^6$ }\\
    \hline
    $ \displaystyle \b{p}\ppi := \f{\p \psi\ppi}{\p \b{\xi}\i} $ & $ \displaystyle  \b{p}\pppi := \int_{\c{H}_0} 2 \, \b{\xi}\ii \times \f{\p \psi\pppinii}{\p \b{a}\i\n\ii}  \, \d V\ii  $ &  $ \displaystyle \b{p}\ppppi := \int_{\c{H}_0} \int_{\c{H}_0} 3 \, \b{\xi}\ii \times \b{\xi}\iii \f{\p \psi\ppppiniiniii}{\p {v}\i\n\ii\n\iii}  \, \d V\iii \, \d V\ii $ \\[6pt]
    \hline\hline
    \multicolumn{3}{|c|}{angular momentum balance}\\
    \hline
    $ \displaystyle \int_{\c{H}_0} \b{\xi}\i \times \b{p}\ppi \, \d V\i \stackrel{!}{=} \bnull $ & $ \displaystyle \int_{\c{H}_0} \b{\xi}\i \times \b{p}\pppi \, \d V\i \stackrel{!}{=} \bnull $ & $ \displaystyle \int_{\c{H}_0} \b{\xi}\i \times \b{p}\ppppi \, \d V\i \stackrel{!}{=} \bnull $ \\[6pt]
    \hline \hline
    \multicolumn{3}{|c|}{suitable deformation measures}\\
    \hline
    $ \displaystyle \oo{l} := |\b{\xi}\i|$ \quad , \quad $ \displaystyle \oo{L} := |\b{\Xi}\i|$ & $ \displaystyle \oo{a} := |\b{a}\i\n\ii| $ \quad , \quad $ \displaystyle \oo{A} := |\b{A}\i\n\ii| $ & $ \displaystyle \oo{v} := |{v}\i\n\ii\n\iii|$ \quad , \quad $ \displaystyle \oo{V} := |{V}\i\n\ii\n\iii|$ \\[6pt]
    \hline \hline
    \multicolumn{3}{|c|}{generic examples of interaction energy densities}\\
    \hline
    $ \displaystyle \psi\ppi= \psi\pp(\oo{l};\oo{L}) $ & $ \displaystyle \psi\pppinii = \psi\ppp(\oo{a};\oo{A})$ & $\displaystyle \psi\ppppiniiniii = \psi\pppp(\oo{v};\oo{V})$ \\[6pt]
    \hline \hline
    \multicolumn{3}{|c|}{specific examples of interaction energy densities}\\
    \hline
    $ \displaystyle \psi\ppi = \frac{1}{2} \, C_1 \, \oo{L} \, \lb \frac{\oo{l}}{\oo{L}} - 1 \rb^2$ & $ \displaystyle \psi\pppinii = \frac{1}{2} \, C_2 \, \oo{A} \, \lb \frac{\oo{a}}{\oo{A}} - 1 \rb^2$ & $ \displaystyle \psi\ppppiniiniii = \frac{1}{2} \, C_3 \, \oo{V} \, \lb \frac{\oo{v}}{\oo{V}} - 1 \rb^2$ \\[6pt]
    \hline
    $ \displaystyle \f{\p \psi\ppi}{\p \b{\xi}\i} = C_1 \, \lb \frac{\oo{l}}{\oo{L}} - 1 \rb \, \frac{\b{\xi}\i}{|\b{\xi}\i|}$ & $ \displaystyle \f{\p \psi\pppinii}{\p \b{a}\i\n\ii} = C_2 \, \lb \frac{\oo{a}}{\oo{A}} - 1 \rb \, \frac{\b{a}\i\n\ii}{|\b{a}\i\n\ii|}$ & $ \displaystyle \f{\p \psi\ppppiniiniii}{\p {v}\i\n\ii\n\iii} = C_3 \, \lb \frac{\oo{v}}{\oo{V}} - 1 \rb \, \frac{v\i\n\ii\n\iii}{|v\i\n\ii\n\iii|}$ \\[6pt]
    \hline
\end{tabular}
\label{tab:BOAM-conseq}
\end{table}

\section{Computational implementation}\label{sec:implementation}

The computational implementation of CPD comprises three major steps.
We begin by replacing the integral equations in Section~\ref{sec:comp-i} with quadrature relations using appropriate weighting coefficients.
Next, in Section~\ref{sec:comp-ii}, a discretised form of the linear momentum balance~(\ref{eq:TD_BOLM_LOC}) is derived.
The discretised balance is a non-linear system of coupled equations of the form $\v{R}=\v{O}$ that is solved using a Newton--Raphson scheme.
To do so, we compute the tangent matrix $\v{K}$ defined as the linearisation of the residual vector $\v{R}$.
Finally, the force densities and their derivatives contributing to $\v{R}$ and $\v{K}$, respectively, are calculated from the constitutive laws associated with the stored energy density given in Section~\ref{sec:comp-iii}.

\begin{rmk}
The proposed computational framework corresponds to the three-dimensional setting.
Nevertheless, both plane-strain and plane-stress assumptions can be recovered via applying appropriate boundary conditions on a 3D domain.
Note that neither ``stress'' nor ``strain'' is present in the peridynamic formulation; they can only be computed through post-processing.
Therefore, the notions of ``plane strain'' or ``plane stress'' become naturally less relevant since they are defined from a local view of continuum mechanics.
Furthermore, it is straightforward to infer a fully two-dimensional counterpart from the provided discussion.
Our formulation in 2D, however, corresponds to a purely two-dimensional case wherein both deformations and forces are absent in the third direction.
This compares to the surface elasticity theory of Gurtin and Murdoch~\cite{Gurtin1975,Javili2010b,Javili2013,Javili2014c}.
Obviously, three-neighbour interactions do not contribute in our 2D formulation.
Both three- and two-dimensional numerical examples are provided in Section~\ref{sec:examples}.
\qed
\end{rmk}

\subsection{From continuous to discretise form}\label{sec:comp-i}

In general. The governing equations of CPD are expressed in integral form.
Unlike CCM, even point-wise equations in CPD include integrals over the horizon.
The points $\c{P}^a$ at which we evaluate the balance of momentum~\eqref{eq:TD_BOLM_LOC} are precisely the \emph{collocation points}.
At each collocation point $\c{P}^a$, we use quadrature rules to evaluate the integrals over the horizon by employing \emph{quadrature points}.
In this contribution, the collocation points coincide identically with the quadrature points, henceforth, we refer to them collectively as \emph{grid points} or simply \emph{points}.
This assumption is made for the sake of simplicity; alternatives will be investigated in a separate contribution.

To proceed, we first discretise the domain by a set of grid points that serve the double-purpose of being collocation and quadrature points.
Every grid point in the present description represents continuum point as opposed to physical particles.
Furthermore, each grid point defines a neighbourhood $\c{H}_0 \subset \c{B}_0$.
Each grid point $\c{P}^{a}$ is identified by its coordinates, $\b{X}^a$ and $\b{x}^a$, in the material and spatial configurations, respectively.
Therefore, the integral of an arbitrary quantity $\sth$ over the domain $\c{B}_0$ is approximated by
\begin{equation}\label{eq:comp-1}
\begin{aligned}
\int_{\c{B}_0} \sth \, \d V = \sum_{a=1}^{\#\c{P}} \sth^a \, V^a \,,
\end{aligned}
\end{equation}
where $V^a$ is the volume of the support domain of the grid point $\c{P}^{a}$ and $\#\c{P}$ is the total number of grid points.
The volume $V^a$ can be computed according to the discretisation strategy employed.
For instance, if the points are chosen based on a Voronoi tessellation, the volume of each Voronoi cell can be assigned to the point $\c{P}^{a}$ at its centre, see~\cite{276-Henke2014}.
It is also relatively straightforward to discretise the domain using the common discretisation tools of the finite element method and then associated each finite element with the point $\c{P}^{a}$ in its barycentre with ${V}^{a}$ equal to the volume of the respective element.
Alternatively, for simple geometries, one can discretise the domain using a uniform grid for which $V^a=\alpha \, \ell^3$ with $\ell$ the grid spacing and $\alpha$ a constant dimensionless correction factor accounting for the size of the support.
If the grid spacing is non-uniform, the parameter $\ell$ can be replaced with a suitable average grid spacing $\ell_{\text{avg}}$.
The summation on the right-hand side of Eq.~(\ref{eq:comp-1}) should correctly represent the volume of the domain itself in the sense that
\begin{equation}\label{eq:comp-2}
\begin{aligned}
\int_{\c{B}_0} \d V = \sum_{a=1}^{\#\c{P}} V^a \,.
\end{aligned}
\end{equation}
Next, we discretise the various multiple integrals that appear in the form
\begin{equation}\label{eq:comp-3}
\begin{aligned}
& \int_{\c{B}_0} \int_{\c{H}_0} \sth \, \d V\i \, \d V \quad , \quad \int_{\c{B}_0} \int_{\c{H}_0} \int_{\c{H}_0} \sth \, \d V\ii \, \d V\i \, \d V \quad , \quad \int_{\c{B}_0} \int_{\c{H}_0} \int_{\c{H}_0} \int_{\c{H}_0} \sth \, \d V\iii \, \d V\ii \, \d V\i \, \d V \, .
\end{aligned}
\end{equation}

Let $\#\c{N}$ denote the number of points within the horizon of the point $\c{P}^a$.
The effective volume of the neighbouring point $i$ contributing to one-neighbour interactions at the point $\c{P}^a$ is denoted as $V_1$, assuming that all neighbours equally contribute.
The effective volumes $V_1$ can be defined by
\begin{equation}\label{eq:comp-5}
\begin{aligned}
	 V_1 := \f{\c{V}_\c{H}}{\#\c{N}_1} \,,
\end{aligned}
\end{equation}
where $\c{V}_\c{H}$ denotes the volume of the neighbourhood of the collocation (continuum) point $\c{P}^{a}$.
For example if the point $\c{P}^{a}$ is completely inside the bulk and entirely surrounded (i.e.\ no part of its horizon extends outside of $\c{B}_0$) then $\c{V}_\c{H}=4/3\,\pi\,\delta_0^3$.
Otherwise, $\c{V}_\c{H}$ is modified by a dimensionless correction factor $\beta<1$, as $\c{V}_\c{H}= \beta \, 4/3\,\pi\,\delta_0^3$ where $\beta$ accounts for the truncated support.
In the definition of the effective volume for one-neighbour interactions~(\ref{eq:comp-5}), the total number of \emph{contributing neighbours} within its horizon is denoted as $\#\c{N}_1$.
A neighbour $\{i\}$ of the point $\c{P}^{a}$ is identified as a \emph{contributing neighbour} if the distance between the pair $\{a,i\}$ is less than or equal to the horizon size of $\delta_0$.
Clearly, all neighbours count as \emph{contributing neighbours} for one-neighbour interactions and thus, $\#\c{N}_1=\#\c{N}$.
As will be made clear, this property does not necessarily hold for two-neighbour and three-neighbour interactions.
The integral~(\ref{eq:comp-3})$_1$ can be formally written in the discretised form as
\begin{equation}\label{eq:comp-6}
\begin{aligned}
\int_{\c{B}_0} \int_{\c{H}_0} \sth \, \d V\i \, \d V = \sum_{a=1}^{\#\c{P}} \sum_{{\substack{i=1 \\ i\neq a}}}^{\#\c{N}} \sth^a_i \, V_1 \, V^a \,,
\end{aligned}
\end{equation}
with $\#\c{N}$ being the total number of neighbours in the neighbourhood of the point $\c{P}^{a}$.
In a similar fashion, the integral~(\ref{eq:comp-3})$_2$ can be formally discretised as
\begin{equation}\label{eq:comp-6-1}
\begin{aligned}
\int_{\c{B}_0} \int_{\c{H}_0} \int_{\c{H}_0} \sth \, \d V\ii \, \d V\i \, \d V = \sum_{a=1}^{\#\c{P}} \sum_{{\substack{i=1 \\ i\neq a}}}^{\#\c{N}} \sum_{{\substack{j=1 \\ j\neq i \\ j\neq a}}}^{\#\c{N}} \sth^a_{ij} \, V_2 \, V^a \,,
\end{aligned}
\end{equation}
where $V_2$ is the the effective volume squared contributing to two-neighbour interactions defined by
\begin{equation}\label{eq:comp-7}
\begin{aligned}
	 V_2 := \f{[\c{V}_\c{H}]^2}{\#\c{N}_2} \,,
\end{aligned}
\end{equation}
with $\#\c{N}_2$ the total number of \emph{contributing pairs} in the neighbourhood of the point $\c{P}^{a}$.

\begin{figure}[h]
    \centering
	\includegraphics[width=0.9\textwidth]{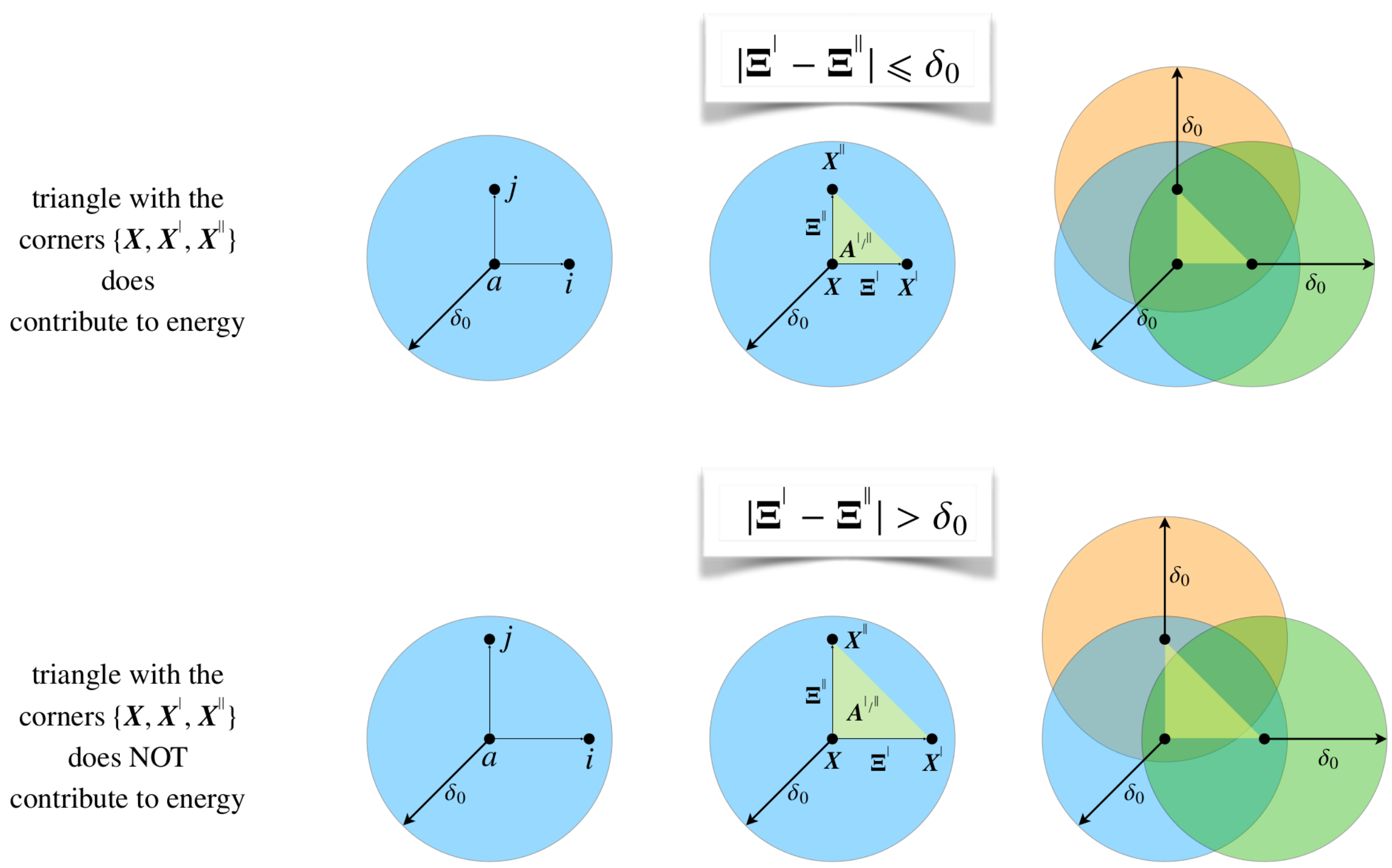}
	\caption{Schematic illustration of how a ``contributing pair'' is defined.
	The pair $i,j$ in the upper half of the figure is a ``contributing pair'' and contributes to the energy due to two-neighbour interactions.
	However, the pair $i,j$ in the lower half is not a ``contributing pair'' and does not contribute to energy.
	}
    \label{fig:binding-pairs}
\end{figure}

A pair of neighbours $\{i,j\}$ of the point $\c{P}^{a}$ is identified as \emph{contributing pair} if (i) the points $\{a,i,j\}$ are non-collinear and (ii) the distance between each pair of $\{a,i\}$, $\{a,j\}$ and $\{i,j\}$ is less than or equal to the horizon size $\delta_0$.
Therefore, and in contrast to one-neighbour interactions, not all the possible pairs would count as contributing pairs, as illustrated in Fig.~\ref{fig:binding-pairs}.
The order of the pairs of neighbours does not matter.
That is, the two neighbour sets $\{i,j\}$ and $\{j,i\}$ contribute equally to the energy.
We exploit this property to improve computational efficiency but omit it in the text, for the sake of readability.

\begin{rmk}
In view of the definition of \emph{contributing pairs}, the first condition must hold since if the points $\{a,i,j\}$ are collinear, the stiffness matrix can become singular.
Furthermore, the derivations of the governing equations of CPD in~\cite{Javili2019} require that if a triangle $\{\overset{\triangle}{aij}\}$ contributes to the energy, both triangles $\{\overset{\triangle}{jai}\}$ and $\{\overset{\triangle}{ija}\}$ must also equally contribute to the energy which leads to the second condition.
That is, if the distance between each pair is not less than or equal to the horizon size $\delta_0$, the stiffness matrix can lose its symmetry.
Similar arguments hold for contributing triplets are defined next.
\qed
\end{rmk}
Finally, the discretised form of the integral~(\ref{eq:comp-3})$_3$ reads
\begin{equation}\label{eq:comp-8}
\begin{aligned}
\int_{\c{B}_0} \int_{\c{H}_0} \int_{\c{H}_0} \int_{\c{H}_0} \sth \, \d V\iii \, \d V\ii \, \d V\i \, \d V = \sum_{a=1}^{\#\c{P}} \sum_{{\substack{i=1 \\ i\neq a}}}^{\#\c{N}} \sum_{{\substack{j=1 \\ j\neq i \\ j\neq a}}}^{\#\c{N}} \sum_{{\substack{k=1 \\ k\neq j \\ k \neq i \\ k \neq a}}}^{\#\c{N}} \sth^a_{ijk} \, V_3 \, V^a  \,,
\end{aligned}
\end{equation}
with
\begin{equation}\label{eq:comp-9}
\begin{aligned}
	 V_3 := \f{[\c{V}_\c{H}]^3}{\#\c{N}_3} \,,
\end{aligned}
\end{equation}
where $V_3$ is the effective volume cubed contributing to three-neighbour interactions in the neighbourhood of the point $\c{P}^{a}$ and $\#\c{N}_3$ is the total number of \emph{contributing triplets} in the neighbourhood.
A triplet of neighbours $\{i,j,k\}$ of the point $\c{P}^{a}$ is identified as \emph{contributing triplet} if (i) the points $\{a,i,j,k\}$ are non-coplanar and (ii) the distance between each pair of $\{a,i\}$, $\{a,j\}$, $\{a,k\}$, $\{i,j\}$ $\{i,k\}$ and $\{k,j\}$ is less than or equal to the horizon size $\delta_0$.
Again, the order of the triplets of neighbours does not matter.
That is, six neighbour sets $\{i,j,k\}$, $\{k,i,j\}$, $\{j,k,i\}$, $\{i,k,j\}$, $\{j,i,k\}$ and $\{k,j,i\}$ contribute equally to the volume.
To test the fidelity of the implementation, one can numerically compute the following integrals to ensure they hold exactly:
\begin{equation}\label{eq:comp-10}
\begin{aligned}
\int_{\c{H}_0} \d V\i = \c{V}_\c{H} \qquad \,, \qquad
\int_{\c{H}_0} \int_{\c{H}_0} \d V\ii \, \d V\i = [\c{V}_\c{H}]^2 \qquad \,, \qquad
\int_{\c{H}_0} \int_{\c{H}_0} \int_{\c{H}_0} \d V\iii \, \d V\ii \, \d V\i  = [\c{V}_\c{H}]^3 \,.
\end{aligned}
\end{equation}

\subsection{Discretised balance of linear momentum}\label{sec:comp-ii}

The underlying governing equation of CPD is the linear momentum balance.
The term $\b{b}\xt_0$ in the linear momentum balance~(\ref{eq:TD_BOLM_LOC}) corresponds to externally prescribed body forces and its incorporation into our framework is fairly straightforward and is omitted from this presentation in order to focus on the novel aspects of the computational implementation.
The point-wise, non-local, form of the linear momentum balance~(\ref{eq:TD_BOLM_LOC}) in the absence of body forces, i.e. the equilibrium equation, is thus given by
\begin{equation}\label{eq:comp-bal-1}
\begin{aligned}
	\int_{\c{H}_0} \b{p}\i \, \d V\i = \bnull \,.
\end{aligned}
\end{equation}
We proceed with this reduced linear momentum balance and develop a discretised version based on the strategy presented in Section~\ref{sec:comp-i}.
Using the definitions of the force densities~(\ref{eq:force-densities}) at each collocation point, the integral over the horizon in~(\ref{eq:comp-bal-1}) can be decomposed into three parts, corresponding to one-, two- and three-neighbour interactions, as follows
\begin{equation}\label{eq:comp-bal-2}
\begin{aligned}
	\int_{\c{H}_0} \b{p}\ppi \, \d V\i + \int_{\c{H}_0} \b{p}\pppi \, \d V\i + \int_{\c{H}_0} \b{p}\ppppi \, \d V\i = \bnull \,.
\end{aligned}
\end{equation}
This form of the point-wise balance of linear momentum can be expressed as
\begin{equation}\label{eq:comp-bal-3}
\begin{aligned}
	 \b{R} = \bnull \qquad \text{with} \qquad \b{R} :=\b{R}\pp + \b{R}\ppp + \b{R}\pppp = \bnull \,,
\end{aligned}
\end{equation}
where $\b{R}$ is the point-wise residual vector and is decomposed into $\b{R}\pp$, $\b{R}\ppp$ and $\b{R}\pppp$ corresponding to one-neighbour, two-neighbour and three-neighbour contributions, respectively.
That is,
\begin{equation}\label{eq:comp-bal-4}
\begin{aligned}
	 & \b{R}\pp := \int_{\c{H}_0} \b{p}\ppi \, \d V\i  = \int_{\c{H}_0} {\f{\p \psi\ppi}{\p \b{\xi}\i}} \, \d V\i  \,, \\
	 & \b{R}\ppp := \int_{\c{H}_0} \b{p}\pppi \, \d V\i   = \int_{\c{H}_0} \int_{\c{H}_0} 2 \, \b{\xi}\ii \times \f{\p \psi\pppinii}{\p \b{a}\i\n\ii} \, \d V\ii \, \d V\i  \,, \\
	 & \b{R}\pppp := \int_{\c{H}_0} \b{p}\ppppi  \, \d V\i = \int_{\c{H}_0} \int_{\c{H}_0} \int_{\c{H}_0} 3 \, \b{\xi}\ii \times \b{\xi}\iii \, \f{\p \psi\ppppiniiniii}{\p v\i\n\ii\n\iii} \, \d V\iii \, \d V\ii \, \d V\i \,.
\end{aligned}
\end{equation}
Next, we discretise the residual vector $\b{R}$.
The \emph{global discretised} residual vector $\v{R}$ is composed of \emph{point-wise discretised} residual vectors $\v{R}^a$ assembled into a global vector as follows
\begin{equation}\label{eq:comp-bal-5}
\v{R} = \left[
\begin{array}{c}
	\v{R}^1 \\ \v{R}^2 \\ \vdots \\ \v{R}^a \\ \vdots \\ \v{R}^{\#\c{P}}
\end{array}\right] = \left[
\begin{array}{c}
	\v{R}\pp^1 + \v{R}\ppp^1 + \v{R}\pppp^1 \\ \v{R}\pp^2 + \v{R}\ppp^2 + \v{R}\pppp^2 \\ \vdots \\ \v{R}\pp^a + \v{R}\ppp^a + \v{R}\pppp^a \\ \vdots \\ \v{R}\pp^{\#\c{P}} + \v{R}\ppp^{\#\c{P}} + \v{R}\pppp^{\#\c{P}}
\end{array}\right] \,.
\end{equation}
The point-wise discretised residual vector $\v{R}^a$ of collocation point $\c{P}^a$ is comprised of the point-wise discretised residual vectors $\v{R}\pp^a$, $\v{R}\ppp^a$ and $\v{R}\pppp^a$ corresponding to one-neighbour, two-neighbour and three-neighbour interactions, respectively.
That is
\begin{tcolorbox}[top=-1mm,bottom=-0mm,ams equation]\label{eq:comp-bal-6}
\begin{aligned}
& \v{R}\pp^a :=\sum_{{\substack{i=1 \\ i\neq a}}}^{\#\c{N}} {\f{\p \psi\ppi}{\p \b{\xi}\i}} \, V_1  \,, \\
& \v{R}\ppp^a := \sum_{{\substack{i=1 \\ i\neq a}}}^{\#\c{N}} \sum_{{\substack{j=1 \\ j\neq i \\ j\neq a}}}^{\#\c{N}} 2 \, \b{\xi}\ii \times \f{\p \psi\pppinii}{\p \b{a}\i\n\ii} \, V_2  \,, \\
& \v{R}\pppp^a := \sum_{{\substack{i=1 \\ i\neq a}}}^{\#\c{N}} \sum_{{\substack{j=1 \\ j\neq i \\ j\neq a}}}^{\#\c{N}} \sum_{{\substack{k=1 \\ k\neq j \\ k \neq i \\ k \neq a}}}^{\#\c{N}} 3 \, \b{\xi}\ii \times \b{\xi}\iii \, \f{\p \psi\ppppiniiniii}{\p v\i\n\ii\n\iii} \, V_3   \,.
\end{aligned}
\end{tcolorbox}
\noindent In the definitions of the point-wise discretised residual vectors~(\ref{eq:comp-bal-6}), the bond vectors are related to the deformation via the relations
\begin{equation}\label{eq:comp-bal-7}
\b{\xi}\i = \v{x}^i - \v{x}^a \quad , \quad \b{\xi}\ii = \v{x}^j - \v{x}^a \quad , \quad \b{\xi}\iii = \v{x}^k - \v{x}^a \,,
\end{equation}
with $\v{x}^a$ being the position vector of the collocation point $\c{P}^a$ in the deformed configuration.
The deformation vector of the neighbours $i$, $j$ and $k$ are denoted by $\v{x}^i$, $\v{x}^j$ and $\v{x}^k$, respectively.
The global residual vector $\v{R}$ is energetically conjugate to the global deformation vector $\v{x}$ that consists of the point-wise deformation vectors $\v{x}^a$, that is
\begin{equation}\label{eq:comp-bal-8}
\v{x} = \left[
\begin{array}{c}
	\v{x}^1 \\ \v{x}^2 \\ \vdots \\ \v{x}^a \\ \vdots \\ \v{x}^{\#\c{P}}
\end{array}\right] \,.
\end{equation}
As it is customary in nonlinear problems involving large deformations, the full deformation history can be divided into \emph{increments}.
The fully discrete nonlinear system of governing equations at each increment can be concisely stated as $\v{R} \stackrel{\cdot}{=} \v{O}$ whose approximate solution is obtained via an \emph{iterative} Newton--Raphson scheme.
The consistent linearization of the resulting system at iteration $k$ reads
\begin{equation}\label{eq:comp-bal-9}
\begin{aligned}
	\v{R}_{k+1} \stackrel{\cdot}{=} \v{O} \qquad \text{with} \qquad \v{R}_{k+1} = \v{R}_{k} + \f{\p \v{R}}{\p \v{x}} \bigg|_{k} \cdot \Delta \v{x}_k  \qquad \Rightarrow \qquad \v{R}_{k} + \f{\p \v{R}}{\p \v{x}} \bigg|_{k} \cdot \Delta \v{x}_k \stackrel{\cdot}{=} \v{O} \,,
\end{aligned}
\end{equation}
with the resulting deformation change $\Delta \v{x}_k$ at iteration $k$ given by
\begin{equation}\label{eq:comp-bal-10}
\begin{aligned}
	\Delta \v{x}_k = - \v{K}\inv_k \cdot \v{R}_{k} \qquad \text{with} \qquad \v{K}_k := \f{\p \v{R}}{\p \v{x}} \bigg|_{k} \,,
\end{aligned}
\end{equation}
where $\v{K}_k$ denotes the corresponding algorithmic tangent (stiffness) at iteration $k$.
Finally, the deformation $\v{x}$ is updated after each iteration by $\Delta \v{x}$ obtained from Eq.~(\ref{eq:comp-bal-10}) according to
\begin{equation}\label{eq:comp-bal-11}
\begin{aligned}
	\v{x}_{k+1} = \v{x}_k + \Delta \v{x}_k \,.
\end{aligned}
\end{equation}
Note that the tangent $\v{K}$ is a matrix and composed of point-wise contributions $\v{K}^{ab}$.
That is
\begin{equation}\label{eq:comp-bal-12}
\v{K} = \left[
\begin{array}{cccccc}
	\v{K}^{11} & \v{K}^{12} & \dots & \v{K}^{1b} & \dots & \v{K}^{1 {\#\c{P}}} \\
	\v{K}^{21} & \v{K}^{22} & \dots & \v{K}^{2b} & \dots & \v{K}^{2 {\#\c{P}}} \\
	\vdots & \vdots & \vdots & \vdots & \vdots & \vdots \\
	\v{K}^{a1} & \v{K}^{a2} & \dots & \v{K}^{ab} & \dots & \v{K}^{a {\#\c{P}}} \\
	\vdots & \vdots & \vdots & \vdots & \vdots & \vdots \\
	\v{K}^{{\#\c{P}}1} & \v{K}^{{\#\c{P}}2} & \dots & \v{K}^{{\#\c{P}}b} & \dots & \v{K}^{{\#\c{P}} {\#\c{P}}} \\
\end{array}\right] \qquad \text{with} \qquad \v{K}^{ab} =  \f{\p \v{R}^a}{\p \v{x}^b} \,.
\end{equation}
Each contribution $\v{K}^{ab}$ itself can be further decomposed into the contributions from one-neighbour, two-neighbour and three-neighbour interactions.
That is
\begin{tcolorbox}[top=-2mm,bottom=2mm,ams equation]\label{eq:comp-bal-13}
\v{K}^{ab} = \v{K}^{ab}_1 + \v{K}^{ab}_2 + \v{K}^{ab}_3 \qquad \text{with} \qquad \v{K}^{ab}_1 = \f{\p \v{R}^a_1}{\p \v{x}^b} \quad , \quad \v{K}^{ab}_2 = \f{\p \v{R}^a_2}{\p \v{x}^b} \quad , \quad \v{K}^{ab}_3 = \f{\p \v{R}^a_3}{\p \v{x}^b}  \,.
\end{tcolorbox}
The next task is to compute the discretised point-wise residual $\v{R}^a$ and the discretised point-wise stiffness $\v{K}^{ab}$ from the stored energy densities~(\ref{eq:angmom-6}), (\ref{eq:angmom-7}) and (\ref{eq:angmom-8}) corresponding to one-neighbour, two-neighbour and three-neighbour interactions, respectively.

\subsection{Discretised residuals and tangents}\label{sec:comp-iii}

The final steps in the computational implementation of the proposed scheme are (i) to express the discretised residual vectors~(\ref{eq:comp-bal-6}) in terms of the deformation of the point $\c{P}^{a}$ and its neighbours, and (ii) to compute their associated tangents~(\ref{eq:comp-bal-13}).
We begin with the residual vectors.
Before proceeding, recall the definitions
\begin{equation}\label{eq:comp-tang-1}
\begin{aligned}
& \b{\Xi}\i := \v{X}^i - \v{X}^a && \quad , \quad && \b{\Xi}\ii := \v{X}^j - \v{X}^a && \quad , \quad && \b{\Xi}\iii := \v{X}^k - \v{X}^a \,, \\
& \b{\xi}\i := \v{x}^i - \v{x}^a && \quad , \quad && \b{\xi}\ii := \v{x}^j - \v{x}^a && \quad , \quad && \b{\xi}\iii := \v{x}^k - \v{x}^a \,,
\end{aligned}
\end{equation}
in the material and spatial configuration, respectively.
Furthermore, the following relations will prove useful throughout the forthcoming derivations:
\begin{equation}\label{eq:comp-tang-2}
\begin{aligned}
& \f{\p S_1}{\p \b{\xi}\i} = \f{\p}{\p \b{\xi}\i} \lrp{ \f{\oo{l}}{\oo{L}} } = \f{1}{\oo{L}} \, \f{\p}{\p \b{\xi}\i} \lrp{ |\b{\xi}\i| } = \f{1}{\oo{L}} \, \f{\b{\xi}\i}{|\b{\xi}\i|} \,, \\
& \f{\p S_2}{\p \b{a}\i\n\ii} = \f{\p}{\p \b{a}\i\n\ii} \lrp{ \f{\oo{a}}{\oo{A}} } = \f{1}{\oo{A}} \, \f{\p}{\p \b{a}\i\n\ii} \lrp{ |\b{a}\i\n\ii| } = \f{1}{\oo{A}} \, \f{\b{a}\i\n\ii}{|\b{a}\i\n\ii|} \,, \\
& \f{\p S_3}{\p {v}\i\n\ii\n\iii} = \f{\p}{\p {v}\i\n\ii\n\iii} \lrp{ \f{\oo{v}}{\oo{V}} } = \f{1}{\oo{V}} \, \f{\p}{\p {v}\i\n\ii\n\iii} \lrp{ |{v}\i\n\ii\n\iii| } = \f{1}{\oo{V}} \, \f{{v}\i\n\ii\n\iii}{|\b{v}\i\n\ii\n\iii|} \,.
\end{aligned}
\end{equation}
The point-wise discretised residual vector of point $\c{P}^a$ due to one-neighbour interactions reads
\begin{equation}\label{eq:comp-tang-3}
\begin{aligned}
& \v{R}\pp^a = \sum_{{\substack{i=1 \\ i\neq a}}}^{\#\c{N}} \f{\p \psi\ppi}{\p \b{\xi}\i} \, V_1  \,,
\end{aligned}
\end{equation}
where ${\p \psi\ppi}/{\p \b{\xi}\i}$ can be expressed as
\begin{align}\label{eq:comp-tang-4}
\f{\p \psi\ppi}{\p \b{\xi}\i} &= \f{\p }{\p \b{\xi}\i} \lrp{ \frac{1}{2} \, C_1 \, \oo{L} \, \lb S_1 - 1 \rb^2 } = C_1 \, \oo{L} \, \lrb{ S_1 - 1 }  \, \f{\p S_1}{\p \b{\xi}\i} \,. \nonumber
\intertext{Upon using Eq.~(\ref{eq:comp-tang-2})${}_1$ and the relation $S_1 = {\oo{l}}/{\oo{L}} = {|\b{\xi}\i|}/{|\b{\Xi}\i|}$, one obtains}
\f{\p \psi\ppi}{\p \b{\xi}\i} &= C_1 \, \lrb{ S_1 - 1 }  \, \f{\b{\xi}\i}{|\b{\xi}\i|} = C_1 \, \lrb{ \f{1}{|\b{\Xi}\i|} - \f{1}{|\b{\xi}\i|} } \, {\b{\xi}\i} \,.
\end{align}
Thus, the point-wise discretised residual vector of point $\c{P}^a$ due to one-neighbour interactions reads
\begin{tcolorbox}[top=-1mm,bottom=-0mm,ams equation]\label{eq:comp-tang-5}
\begin{aligned}
& \v{R}\pp^a = \sum_{{\substack{i=1 \\ i\neq a}}}^{\#\c{N}} C_1 \, \lrb{ \f{1}{|\b{\Xi}\i|} - \f{1}{|\b{\xi}\i|} }  \, {\b{\xi}\i} \, V_1 \,.
\end{aligned}
\end{tcolorbox}
\noindent The point-wise discretised residual vector of point $\c{P}^a$ due to two-neighbour interactions is given by
\begin{equation}\label{eq:comp-tang-6}
\begin{aligned}
& \v{R}\ppp^a = \sum_{{\substack{i=1 \\ i\neq a}}}^{\#\c{N}} \sum_{{\substack{j=1 \\ j\neq i \\ j\neq a}}}^{\#\c{N}} 2 \, \b{\xi}\ii \times \f{\p \psi\pppi}{\p \b{a}\i\n\ii}  \, V_2 \,, \\
\end{aligned}
\end{equation}
where ${\p \psi\pppi}/{\p \b{a}\i\n\ii}$ reads
\begin{align}
\f{\p \psi\pppi}{\p \b{a}\i\n\ii} &= \f{\p }{\p \b{a}\i\n\ii} \lrp{ \frac{1}{2} \, C_2 \, \oo{A} \, \lb S_2 - 1 \rb^2 } = C_2 \, \oo{A} \, \lrb{ S_2 - 1 }  \, \f{\p S_2}{\p \b{a}\i\n\ii} \,, \nonumber
\intertext{and using Eq.~(\ref{eq:comp-tang-2})${}_2$ and the relation $S_2 = {\oo{a}}/{\oo{A}} = {|\b{a}\i\n\ii|}/{|\b{A}\i\n\ii|} = {|\b{\xi}\i \times \b{\xi}\ii|}/{|\b{\Xi}\i \times \b{\Xi}\ii|}$ furnishes}
\f{\p \psi\pppi}{\p \b{a}\i\n\ii} &= C_2 \, \lrb{ S_2 - 1 }  \, \f{\b{a}\i\n\ii}{|\b{a}\i\n\ii|} = C_2 \, \lrb{ \f{1}{|\b{A}\i\n\ii|} - \f{1}{|\b{a}\i\n\ii|} }  \, {\b{a}\i\n\ii}
= C_2 \, \lrb{ \f{1}{|\b{\Xi}\i \times \b{\Xi}\ii|} - \f{1}{|\b{\xi}\i \times \b{\xi}\ii|} }  \, {\b{\xi}\i \times \b{\xi}\ii} \label{eq:comp-tang-7} \,.
\end{align}
Inserting Eq.~(\ref{eq:comp-tang-7}) into Eq.~(\ref{eq:comp-tang-6}) yields
\begin{equation}\label{eq:comp-tang-8}
\begin{aligned}
& \v{R}\ppp^a = \sum_{{\substack{i=1 \\ i\neq a}}}^{\#\c{N}} \sum_{{\substack{j=1 \\ j\neq i \\ j\neq a}}}^{\#\c{N}} 2 \, C_2 \, \lrb{ \f{1}{|\b{\Xi}\i \times \b{\Xi}\ii|} - \f{1}{|\b{\xi}\i \times \b{\xi}\ii|} }  \, \b{\xi}\ii \times {\b{\xi}\i \times \b{\xi}\ii}  \, V_2  \,. \\
\end{aligned}
\end{equation}
Using the identity
\begin{equation}\label{eq:comp-tang-9}
\begin{aligned}
	\b{\xi}\ii \times \b{\xi}\i \times \b{\xi}\ii = [\b{\xi}\ii \cdot \b{\xi}\ii] \, \b{\xi}\i - [\b{\xi}\ii \cdot \b{\xi}\i] \, \b{\xi}\ii \,,
\end{aligned}
\end{equation}
the point-wise discretised residual vector of point $\c{P}^a$ due to two-neighbour interactions reads
\begin{tcolorbox}[top=-1mm,bottom=-0mm,ams equation]\label{eq:comp-tang-10}
\begin{aligned}
& \v{R}\ppp^a = \sum_{{\substack{i=1 \\ i\neq a}}}^{\#\c{N}} \sum_{{\substack{j=1 \\ j\neq i \\ j\neq a}}}^{\#\c{N}} 2 \, C_2 \, \lrb{ \f{1}{|\b{\Xi}\i \times \b{\Xi}\ii|} - \f{1}{|\b{\xi}\i \times \b{\xi}\ii|} }  \, \Big[ [\b{\xi}\ii \cdot \b{\xi}\ii] \, \b{\xi}\i - [\b{\xi}\ii \cdot \b{\xi}\i] \, \b{\xi}\ii \Big] \, V_2 \,.
\end{aligned}
\end{tcolorbox}
\noindent Lastly, the point-wise discretised residual vector of point $\c{P}^a$ due to three-neighbour interactions
\begin{equation}\label{eq:comp-tang-11}
\begin{aligned}
& \v{R}\pppp^a= \sum_{{\substack{i=1 \\ i\neq a}}}^{\#\c{N}} \sum_{{\substack{j=1 \\ j\neq i \\ j\neq a}}}^{\#\c{N}} \sum_{{\substack{k=1 \\ k\neq j \\ k \neq i \\ k \neq a}}}^{\#\c{N}} 3 \, [\b{\xi}\ii \times \b{\xi}\iii] \f{\p \psi\ppppi}{\p {v}\i\n\ii\n\iii} \, V_3   \,,
\end{aligned}
\end{equation}
is rewritten using the relation
\begin{align}
\f{\p \psi\ppppi}{\p {v}\i\n\ii\n\iii} &= \f{\p }{\p {v}\i\n\ii\n\iii} \lrp{ \frac{1}{2} \, C_3 \, \oo{V} \, \lb S_3 - 1 \rb^2 } = C_3 \, \oo{V} \, \lrb{ S_3 - 1 }  \, \f{\p S_3}{\p {v}\i\n\ii\n\iii} \,, \nonumber \\
\intertext{and using Eq.~(\ref{eq:comp-tang-2})${}_3$ and the relation $S_3 = {\oo{v}}/{\oo{V}} = |{v}\i\n\ii\n\iii|/|{V}\i\n\ii\n\iii| = {|[\b{\xi}\i \times \b{\xi}\ii]\cdot\b{\xi}\iii|}/{|[\b{\Xi}\i \times \b{\Xi}\ii]\cdot\b{\Xi}\iii|}$ gives}
\f{\p \psi\ppppi}{\p {v}\i\n\ii\n\iii} &= C_3 \, \lrb{ S_3 - 1 }  \, \f{{v}\i\n\ii\n\iii}{|{v}\i\n\ii\n\iii|} = C_3 \, \lrb{ \f{1}{|{V}\i\n\ii\n\iii|} - \f{1}{|{v}\i\n\ii\n\iii|} }  \, {{v}\i\n\ii\n\iii} \nonumber \\[9pt]
& = C_3 \, \lrb{ \f{1}{\big|[\b{\Xi}\i \times \b{\Xi}\ii]\cdot\b{\Xi}\iii\big|} - \f{1}{\big|[\b{\xi}\i \times \b{\xi}\ii]\cdot\b{\xi}\iii\big|} } \, \lrb{[\b{\xi}\i \times \b{\xi}\ii]\cdot\b{\xi}\iii} \label{eq:comp-tang-12} \,.
\end{align}
Inserting Eq.~(\ref{eq:comp-tang-12}) into Eq.~(\ref{eq:comp-tang-11}) yields the point-wise discretised residual vector of point $\c{P}^a$ due to three-neighbour interactions as
\begin{tcolorbox}[top=-1mm,bottom=-0mm,ams equation]\label{eq:comp-tang-13}
\begin{aligned}
& \v{R}\pppp^a= \sum_{{\substack{i=1 \\ i\neq a}}}^{\#\c{N}} \sum_{{\substack{j=1 \\ j\neq i \\ j\neq a}}}^{\#\c{N}} \sum_{{\substack{k=1 \\ k\neq j \\ k \neq i \\ k \neq a}}}^{\#\c{N}} 3 \, C_3 \, [\b{\xi}\ii \times \b{\xi}\iii] \lrb{ \f{1}{\big|[\b{\Xi}\i \times \b{\Xi}\ii]\cdot\b{\Xi}\iii\big|} - \f{1}{\big|[\b{\xi}\i \times \b{\xi}\ii]\cdot\b{\xi}\iii\big|} } \, \lrb{[\b{\xi}\i \times \b{\xi}\ii]\cdot\b{\xi}\iii}  \, V_3 \,.
\end{aligned}
\end{tcolorbox}

Equipped with the discretised residuals~(\ref{eq:comp-tang-5}), (\ref{eq:comp-tang-10}) and (\ref{eq:comp-tang-13}), the algorithmic tangents are derived next.
Unlike the commonly accepted strategy in classical state-based peridynamics, we do not approximate the tangent stiffness using finite difference or other numerical differentiation schemes.
A key feature of the proposed methodology is that we compute the tangent stiffness $\v{K}$ directly.
This is mainly possible since we do not rely on the notion of ``state''.
Computing $\v{K}$ directly has enormous advantages.
For example, the associated decrease in the number of Newton iterations required to achieve convergence can reduce the computational time and significantly boost the accuracy of the calculations.
Throughout our numerical simulations we observe the asymptotic quadratic convergence associated with the Newton--Raphson scheme.
Note, for highly non-linear sets of equations, numerical differentiation is inaccurate and ultimately unstable \citep{Rudraraju2014}.
We derive the tangents for pairs of points $\c{P}^a$ and $\c{P}^b$ due to one-neighbour, two-neighbour and three-neighbour interactions separately and combine them additively according to Eq.~(\ref{eq:comp-bal-13}).
The discretised tangent matrix for the points $\c{P}^a$ and $\c{P}^b$ due to one-neighbour interactions reads
\begin{equation}\label{eq:comp-tang-14}
\begin{aligned}
	\v{K}^{ab}_1 = \f{\p \v{R}^a_1}{\p \v{x}^b} \,,
\end{aligned}
\end{equation}
which after using Eq.~(\ref{eq:comp-tang-5}), can be written as
\begin{equation}\label{eq:comp-tang-15}
\begin{aligned}
	\f{\p \v{R}^a_1}{\p \v{x}^b} &= \f{\p }{\p \v{x}^b} \lrp{ \sum_{{\substack{i=1 \\ i\neq a}}}^{\#\c{N}} C_1 \, \lrb{ \f{1}{|\b{\Xi}\i|} - \f{1}{|\b{\xi}\i|} } \, {\b{\xi}\i}  \, V_1   } \\
	&= \sum_{{\substack{i=1 \\ i\neq a}}}^{\#\c{N}} C_1 \, \f{\p }{\p \v{x}^b} \lrp{ \lrb{ \f{1}{|\b{\Xi}\i|} - \f{1}{|\b{\xi}\i|} } \, {\b{\xi}\i} } \, V_1  \,.
\end{aligned}
\end{equation}
To proceed, we use the chain rule
\begin{equation}\label{eq:comp-tang-15-2}
\begin{aligned}
	\f{\p \sth }{\p \v{x}^b} &= \f{\p \sth }{\p \b{\xi}\i} \cdot \f{\p \b{\xi}\i }{\p \v{x}^b}  \\
	&= \f{\p \sth }{\p \b{\xi}\i} \cdot \bigg[ \lrb{ \delta^{ib} - \delta^{ab} } \, \b{i} \bigg]  \\
	&= \lrb{ \delta^{ib} - \delta^{ab} } \, \f{\p \sth }{\p \b{\xi}\i} \,,
\end{aligned}
\end{equation}
and the relation
\begin{equation}\label{eq:comp-tang-15-3}
\begin{aligned}
	\f{\p }{\p \b{\xi}\i} \lrp{ \lrb{ \f{1}{|\b{\Xi}\i|} - \f{1}{|\b{\xi}\i|} } \, {\b{\xi}\i} } = \f{1}{|\b{\xi}\i|^3} \, {\b{\xi}\i} \dyad {\b{\xi}\i} + \lrb{ \f{1}{|\b{\Xi}\i|} - \f{1}{|\b{\xi}\i|} } \b{i} \,,
\end{aligned}
\end{equation}
where $\b{i}$ is the identity tensor.
Thus, the discretised tangent for the points $\c{P}^a$ and $\c{P}^b$ due to one-neighbour interactions reads
\begin{tcolorbox}[top=-1mm,bottom=-0mm,ams equation]\label{eq:comp-tang-16}
\begin{aligned}
& \v{K}^{ab}_1 = \f{\p \v{R}^a_1}{\p \v{x}^b} = \sum_{{\substack{i=1 \\ i\neq a}}}^{\#\c{N}} C_1 \, \lrb{ \delta^{ib} - \delta^{ab} } \, \lrb{\, \f{1}{|\b{\xi}\i|^3} \, {\b{\xi}\i} \dyad {\b{\xi}\i} + \lrb{ \f{1}{|\b{\Xi}\i|} - \f{1}{|\b{\xi}\i|} } \b{i} \,} V_1  \,.
\end{aligned}
\end{tcolorbox}
\noindent The discretised tangent matrix for the points $\c{P}^a$ and $\c{P}^b$ due to two-neighbour interactions is
\begin{equation}\label{eq:comp-tang-17}
\begin{aligned}
	\v{K}^{ab}_2 = \f{\p \v{R}^a_2}{\p \v{x}^b} \,,
\end{aligned}
\end{equation}
which, after using Eq.~(\ref{eq:comp-tang-11}), can be written as
\begin{equation}\label{eq:comp-tang-18}
\begin{aligned}
	\f{\p \v{R}^a_2}{\p \v{x}^b} &= \f{\p }{\p \v{x}^b} \lrp{ \sum_{{\substack{i=1 \\ i\neq a}}}^{\#\c{N}} \sum_{{\substack{j=1 \\ j\neq i \\ j\neq a}}}^{\#\c{N}} 2 \, C_2 \, \lrb{ \f{1}{|\b{\Xi}\i \times \b{\Xi}\ii|} - \f{1}{|\b{\xi}\i \times \b{\xi}\ii|} }  \, \Big[ [\b{\xi}\ii \cdot \b{\xi}\ii] \, \b{\xi}\i - [\b{\xi}\ii \cdot \b{\xi}\i] \, \b{\xi}\ii \Big] \, V_2  } \\
	&= \sum_{{\substack{i=1 \\ i\neq a}}}^{\#\c{N}} \sum_{{\substack{j=1 \\ j\neq i \\ j\neq a}}}^{\#\c{N}} 2 \, C_2 \, \f{\p }{\p \v{x}^b} \lrp{ \, \lrb{ \f{1}{|\b{\Xi}\i \times \b{\Xi}\ii|} - \f{1}{|\b{\xi}\i \times \b{\xi}\ii|} }  \, \Big[ [\b{\xi}\ii \cdot \b{\xi}\ii] \, \b{\xi}\i - [\b{\xi}\ii \cdot \b{\xi}\i] \, \b{\xi}\ii \Big]
	\, } \, V_2  \,.
\end{aligned}
\end{equation}
To proceed, we use the chain rule
\begin{equation}\label{eq:comp-tang-19}
\begin{aligned}
	\f{\p \sth }{\p \v{x}^b} &= \f{\p \sth }{\p \b{\xi}\i} \cdot \f{\p \b{\xi}\i }{\p \v{x}^b} + \f{\p \sth }{\p \b{\xi}\ii} \cdot \f{\p \b{\xi}\ii }{\p \v{x}^b} \\
	&= \f{\p \sth }{\p \b{\xi}\i} \cdot \bigg[ \lrb{ \delta^{ib} - \delta^{ab} } \, \b{i} \bigg] + \f{\p \sth }{\p \b{\xi}\ii} \cdot \bigg[ \lrb{ \delta^{jb} - \delta^{ab} } \, \b{i} \bigg] \\
	&= \lrb{ \delta^{ib} - \delta^{ab} } \, \f{\p \sth }{\p \b{\xi}\i} + \lrb{ \delta^{jb} - \delta^{ab} } \, \f{\p \sth }{\p \b{\xi}\ii} \,,
\end{aligned}
\end{equation}
and the identities
\begin{equation}\label{eq:comp-tang-21}
\begin{aligned}
	& \f{\p }{\p \b{\xi}\i} \lrp{ \f{1}{|\b{\xi}\i \times \b{\xi}\ii|} } = \f{1}{|\b{\xi}\i \times \b{\xi}\ii|^3} \, \Big[ [\b{\xi}\ii \cdot \b{\xi}\i] \, \b{\xi}\ii - [\b{\xi}\ii \cdot \b{\xi}\ii] \, \b{\xi}\i \Big] \,, \\[6pt]
	& \f{\p }{\p \b{\xi}\ii} \lrp{ \f{1}{|\b{\xi}\i \times \b{\xi}\ii|} } = \f{1}{|\b{\xi}\i \times \b{\xi}\ii|^3} \, \Big[ [\b{\xi}\ii \cdot \b{\xi}\i] \, \b{\xi}\i - [\b{\xi}\i \cdot \b{\xi}\i] \, \b{\xi}\ii \Big] \,, \\[6pt]
	& \f{\p }{\p \b{\xi}\i} \lrp{ \b{\xi}\ii \cdot \b{\xi}\i \, \b{\xi}\ii - \b{\xi}\ii \cdot \b{\xi}\ii \, \b{\xi}\i } = \b{\xi}\ii \dyad \b{\xi}\ii - [\b{\xi}\ii \cdot \b{\xi}\ii] \, \b{i} \,, \\[6pt]
	& \f{\p }{\p \b{\xi}\ii} \lrp{ \b{\xi}\ii \cdot \b{\xi}\i \, \b{\xi}\ii - \b{\xi}\ii \cdot \b{\xi}\ii \, \b{\xi}\i } = \b{\xi}\ii \dyad \b{\xi}\i + [\b{\xi}\ii \cdot \b{\xi}\i] \, \b{i} - 2 \, \b{\xi}\i \dyad \b{\xi}\ii \,, \\
\end{aligned}
\end{equation}
the derivations of which are omitted for the sake of brevity.
Inserting relations~(\ref{eq:comp-tang-21}) into~(\ref{eq:comp-tang-19}),
the discretised tangent for the points $\c{P}^a$ and $\c{P}^b$ due to two-neighbour interactions reads
\begin{tcolorbox}[top=-1mm,bottom=-0mm,ams equation]\label{eq:comp-tang-23}
\begin{aligned}
\v{K}^{ab}_2 &= \sum_{{\substack{i=1 \\ i\neq a}}}^{\#\c{N}} \sum_{{\substack{j=1 \\ j\neq i \\ j\neq a}}}^{\#\c{N}} 2 \, C_2 \, \lrb{ \delta^{ib} - \delta^{ab} } \, \f{1}{|\b{\xi}\i \times \b{\xi}\ii|^3}  \, \lrb{ [\b{\xi}\ii \cdot \b{\xi}\i] \, \b{\xi}\ii - [\b{\xi}\ii \cdot \b{\xi}\ii] \, \b{\xi}\i } \dyad \lrb{ [\b{\xi}\ii \cdot \b{\xi}\i] \, \b{\xi}\ii - [\b{\xi}\ii \cdot \b{\xi}\ii] \, \b{\xi}\i } \, V_2  \\
& + \sum_{{\substack{i=1 \\ i\neq a}}}^{\#\c{N}} \sum_{{\substack{j=1 \\ j\neq i \\ j\neq a}}}^{\#\c{N}} 2 \, C_2 \, \lrb{ \delta^{ib} - \delta^{ab} } \, \lrb{ \f{1}{|\b{\xi}\i \times \b{\xi}\ii|} - \f{1}{|\b{\Xi}\i \times \b{\Xi}\ii|} } \, \lrb{ \b{\xi}\ii \dyad \b{\xi}\ii - [\b{\xi}\ii \cdot \b{\xi}\ii] \, \b{i} } \, V_2  \\
& + \sum_{{\substack{i=1 \\ i\neq a}}}^{\#\c{N}} \sum_{{\substack{j=1 \\ j\neq i \\ j\neq a}}}^{\#\c{N}} 2 \, C_2 \, \lrb{ \delta^{jb} - \delta^{ab} } \, \f{1}{|\b{\xi}\i \times \b{\xi}\ii|^3} \, \lrb{ [\b{\xi}\ii \cdot \b{\xi}\i] \, \b{\xi}\ii - [\b{\xi}\ii \cdot \b{\xi}\ii] \, \b{\xi}\i } \dyad \lrb{ [\b{\xi}\i \cdot \b{\xi}\ii] \, \b{\xi}\i - [\b{\xi}\i \cdot \b{\xi}\i] \, \b{\xi}\ii } \, V_2  \\
& + \sum_{{\substack{i=1 \\ i\neq a}}}^{\#\c{N}} \sum_{{\substack{j=1 \\ j\neq i \\ j\neq a}}}^{\#\c{N}} 2 \, C_2 \, \lrb{ \delta^{jb} - \delta^{ab} } \, \lrb{ \f{1}{|\b{\xi}\i \times \b{\xi}\ii|} - \f{1}{|\b{\Xi}\i \times \b{\Xi}\ii|} }  \, \lrb{ \b{\xi}\ii \dyad \b{\xi}\i + [\b{\xi}\ii \cdot \b{\xi}\i] \, \b{i} - 2 \, \b{\xi}\i \dyad \b{\xi}\ii  } \, V_2  \,.
\end{aligned}
\end{tcolorbox}
\noindent One may attempt to rewrite the tangent~(\ref{eq:comp-tang-23}) in a more compact form since similar terms appear on different lines.
However, for the sake of clarity, we retain this expanded version as it immediately follows from the previous derivations.
Finally, the discretised tangent matrix for the points $\c{P}^a$ and $\c{P}^b$ due to three-neighbour interactions
\begin{equation}\label{eq:comp-tang-24}
\begin{aligned}
	\v{K}^{ab}_3 = \f{\p \v{R}^a_3}{\p \v{x}^b} \,,
\end{aligned}
\end{equation}
can be expressed using the relation
\begin{equation}\label{eq:comp-tang-25}
\begin{aligned}
	\f{\p \v{R}^a_3}{\p \v{x}^b} &= \f{\p }{\p \v{x}^b} \lrp{ \sum_{{\substack{i=1 \\ i\neq a}}}^{\#\c{N}} \sum_{{\substack{j=1 \\ j\neq i \\ j\neq a}}}^{\#\c{N}} \sum_{{\substack{k=1 \\ k\neq j \\ k \neq i \\ k \neq a}}}^{\#\c{N}} 3 \, C_3 \, [\b{\xi}\ii \times \b{\xi}\iii] \lrb{ \f{1}{\big|[\b{\Xi}\i \times \b{\Xi}\ii]\cdot\b{\Xi}\iii\big|} - \f{1}{\big|[\b{\xi}\i \times \b{\xi}\ii]\cdot\b{\xi}\iii\big|} } \, \lrb{[\b{\xi}\i \times \b{\xi}\ii]\cdot\b{\xi}\iii}  \, V_3  } \\[6pt]
	&= \sum_{{\substack{i=1 \\ i\neq a}}}^{\#\c{N}} \sum_{{\substack{j=1 \\ j\neq i \\ j\neq a}}}^{\#\c{N}} \sum_{{\substack{k=1 \\ k\neq j \\ k \neq i \\ k \neq a}}}^{\#\c{N}} 3 \, C_3 \, \f{\p }{\p \v{x}^b} \lrp{ [\b{\xi}\ii \times \b{\xi}\iii] \lrb{ \f{1}{\big|[\b{\Xi}\i \times \b{\Xi}\ii]\cdot\b{\Xi}\iii\big|} - \f{1}{\big|[\b{\xi}\i \times \b{\xi}\ii]\cdot\b{\xi}\iii\big|} } \, \lrb{[\b{\xi}\i \times \b{\xi}\ii]\cdot\b{\xi}\iii} } \, V_3   \,.
\end{aligned}
\end{equation}
To proceed, we once again use the chain rule
\begin{equation}\label{eq:comp-tang-26}
\begin{aligned}
	\f{\p \sth }{\p \v{x}^b} &= \f{\p \sth }{\p \b{\xi}\i} \cdot \f{\p \b{\xi}\i }{\p \v{x}^b} + \f{\p \sth }{\p \b{\xi}\ii} \cdot \f{\p \b{\xi}\ii }{\p \v{x}^b} + \f{\p \sth }{\p \b{\xi}\iii} \cdot \f{\p \b{\xi}\iii }{\p \v{x}^b} \\
	&= \f{\p \sth }{\p \b{\xi}\i} \cdot \bigg[ \lrb{ \delta^{ib} - \delta^{ab} } \, \b{i} \bigg] + \f{\p \sth }{\p \b{\xi}\ii} \cdot \bigg[ \lrb{ \delta^{jb} - \delta^{ab} } \, \b{i} \bigg] + \f{\p \sth }{\p \b{\xi}\iii} \cdot \bigg[ \lrb{ \delta^{kb} - \delta^{ab} } \, \b{i} \bigg] \\
	&= \lrb{ \delta^{ib} - \delta^{ab} } \, \f{\p \sth }{\p \b{\xi}\i} + \lrb{ \delta^{jb} - \delta^{ab} } \, \f{\p \sth }{\p \b{\xi}\ii} + \lrb{ \delta^{kb} - \delta^{ab} } \, \f{\p \sth }{\p \b{\xi}\iii} \,,
\end{aligned}
\end{equation}
and the identities
\begin{equation}\label{eq:comp-tang-27}
\begin{aligned}
	& \f{\p }{\p \b{\xi}\i} \lrp{ \b{\xi}\ii \times \b{\xi}\iii } = \bnull && , && \f{\p }{\p \b{\xi}\i} \lrp{ [\b{\xi}\i \times \b{\xi}\ii]\cdot\b{\xi}\iii] } = \b{\xi}\ii \times \b{\xi}\iii && , && \f{\p }{\p \b{\xi}\i} \lrp{ \f{1}{\big|[\b{\xi}\i \times \b{\xi}\ii]\cdot\b{\xi}\iii\big|} } = - \f{\lrb{[\b{\xi}\i \times \b{\xi}\ii]\cdot\b{\xi}\iii}}{\big|[\b{\xi}\i \times \b{\xi}\ii]\cdot\b{\xi}\iii\big|^3} \, \b{\xi}\ii \times \b{\xi}\iii \,, \\[6pt]
	& \f{\p }{\p \b{\xi}\ii} \lrp{ \b{\xi}\ii \times \b{\xi}\iii } = \b{\varepsilon} \cdot \b{\xi}\iii && , && \f{\p }{\p \b{\xi}\ii} \lrp{ [\b{\xi}\i \times \b{\xi}\ii]\cdot\b{\xi}\iii] } = \b{\xi}\iii \times \b{\xi}\i && , && \f{\p }{\p \b{\xi}\ii} \lrp{ \f{1}{\big|[\b{\xi}\i \times \b{\xi}\ii]\cdot\b{\xi}\iii\big|} } = - \f{\lrb{[\b{\xi}\i \times \b{\xi}\ii]\cdot\b{\xi}\iii}}{\big|[\b{\xi}\i \times \b{\xi}\ii]\cdot\b{\xi}\iii\big|^3} \, \b{\xi}\iii \times \b{\xi}\i  \,, \\[6pt]
	& \f{\p }{\p \b{\xi}\iii} \lrp{ \b{\xi}\ii \times \b{\xi}\iii } = -\b{\varepsilon} \cdot \b{\xi}\ii && , && \f{\p }{\p \b{\xi}\iii} \lrp{ [\b{\xi}\i \times \b{\xi}\ii]\cdot\b{\xi}\iii] } = \b{\xi}\i \times \b{\xi}\ii && , && \f{\p }{\p \b{\xi}\iii} \lrp{ \f{1}{\big|[\b{\xi}\i \times \b{\xi}\ii]\cdot\b{\xi}\iii\big|} } = - \f{\lrb{[\b{\xi}\i \times \b{\xi}\ii]\cdot\b{\xi}\iii}}{\big|[\b{\xi}\i \times \b{\xi}\ii]\cdot\b{\xi}\iii\big|^3} \, \b{\xi}\i \times \b{\xi}\ii  \,,
\end{aligned}
\end{equation}
the derivations of which are omitted for the sake of brevity.
Inserting relations~(\ref{eq:comp-tang-27}) into~(\ref{eq:comp-tang-25}),
the discretised tangent for the points $\c{P}^a$ and $\c{P}^b$ due to three-neighbour interactions reads
\begin{tcolorbox}[top=-1mm,bottom=-0mm,ams equation]\label{eq:comp-tang-28}
\begin{aligned}
\v{K}^{ab}_3 &= \sum_{{\substack{i=1 \\ i\neq a}}}^{\#\c{N}} \sum_{{\substack{j=1 \\ j\neq i \\ j\neq a}}}^{\#\c{N}} \sum_{{\substack{k=1 \\ k\neq j \\ k \neq i \\ k \neq a}}}^{\#\c{N}} 3 \, C_3 \, \lrb{ \delta^{ib} - \delta^{ab} } \, \f{1}{\big|[\b{\Xi}\i \times \b{\Xi}\ii]\cdot\b{\Xi}\iii\big|} \, \lrb{ [\b{\xi}\ii \times \b{\xi}\iii] \dyad [\b{\xi}\ii \times \b{\xi}\iii] } \, V_3  \\
& + \sum_{{\substack{i=1 \\ i\neq a}}}^{\#\c{N}} \sum_{{\substack{j=1 \\ j\neq i \\ j\neq a}}}^{\#\c{N}} \sum_{{\substack{k=1 \\ k\neq j \\ k \neq i \\ k \neq a}}}^{\#\c{N}} 3 \, C_3 \, \lrb{ \delta^{jb} - \delta^{ab} } \, \lrb{ \f{1}{\big|[\b{\Xi}\i \times \b{\Xi}\ii]\cdot\b{\Xi}\iii\big|} - \f{1}{\big|[\b{\xi}\i \times \b{\xi}\ii]\cdot\b{\xi}\iii\big|} } \, \lrb{[\b{\xi}\i \times \b{\xi}\ii]\cdot\b{\xi}\iii} \, \lrb{\b{\varepsilon} \cdot \b{\xi}\iii} \, V_3  \\
& + \sum_{{\substack{i=1 \\ i\neq a}}}^{\#\c{N}} \sum_{{\substack{j=1 \\ j\neq i \\ j\neq a}}}^{\#\c{N}} \sum_{{\substack{k=1 \\ k\neq j \\ k \neq i \\ k \neq a}}}^{\#\c{N}} 3 \, C_3 \, \lrb{ \delta^{jb} - \delta^{ab} } \, \f{1}{\big|[\b{\Xi}\i \times \b{\Xi}\ii]\cdot\b{\Xi}\iii\big|} \lrb{ [\b{\xi}\ii \times \b{\xi}\iii] \dyad [\b{\xi}\iii \times \b{\xi}\i] } \, V_3  \\
& - \sum_{{\substack{i=1 \\ i\neq a}}}^{\#\c{N}} \sum_{{\substack{j=1 \\ j\neq i \\ j\neq a}}}^{\#\c{N}} \sum_{{\substack{k=1 \\ k\neq j \\ k \neq i \\ k \neq a}}}^{\#\c{N}} 3 \, C_3 \, \lrb{ \delta^{kb} - \delta^{ab} } \, \lrb{ \f{1}{\big|[\b{\Xi}\i \times \b{\Xi}\ii]\cdot\b{\Xi}\iii\big|} - \f{1}{\big|[\b{\xi}\i \times \b{\xi}\ii]\cdot\b{\xi}\iii\big|} } \, \lrb{[\b{\xi}\i \times \b{\xi}\ii]\cdot\b{\xi}\iii} \, \lrb{\b{\varepsilon} \cdot \b{\xi}\ii} \, V_3  \\
& + \sum_{{\substack{i=1 \\ i\neq a}}}^{\#\c{N}} \sum_{{\substack{j=1 \\ j\neq i \\ j\neq a}}}^{\#\c{N}} \sum_{{\substack{k=1 \\ k\neq j \\ k \neq i \\ k \neq a}}}^{\#\c{N}} 3 \, C_3 \, \lrb{ \delta^{kb} - \delta^{ab} } \, \f{1}{\big|[\b{\Xi}\i \times \b{\Xi}\ii]\cdot\b{\Xi}\iii\big|} \lrb{ [\b{\xi}\ii \times \b{\xi}\iii] \dyad [\b{\xi}\i \times \b{\xi}\ii] } \, V_3 
 \,.
\end{aligned}
\end{tcolorbox}
\noindent Again, since similar terms appear on different lines of the tangent~(\ref{eq:comp-tang-28}), it could be written in a more compact form.
However, for the sake of clarity, we retain this expanded version.

The stiffness components due to one-neighbour~(\ref{eq:comp-tang-16}), two-neighbour~(\ref{eq:comp-tang-23}) and three-neighbour~(\ref{eq:comp-tang-28}) interactions, are all symmetric for the variationally consistent mechanical problem of interest here wherein the residuals derive from a potential.
While in some cases, such as the first and second lines of Eq.~(\ref{eq:comp-tang-23}), the symmetric structure of the stiffness is obvious, in other cases, such as the third and fourth lines of Eq.~(\ref{eq:comp-tang-23}) the symmetry is not evident.
The reason for this is that, for instance, the term $\b{\xi}\i \dyad \b{\xi}\ii$ is calculated twice for any given set of $\{i^\star,j^\star\}$ for which $i=i^\star$ and $j=j^\star$ for the first time but $i=j^\star$ and $j=i^\star$ for the second time.
This property could indeed allow us to reduce the number of loops over the neighbours and  avoid multiple counting.
For instance, the index $j$ instead of counting from 1 could start from $i+1$ if the associated implications are accounted for.
However, for the sake of brevity, we do not include these additional steps in the presentation.
This procedure is essentially a technical programming detail also relevant in molecular dynamics simulations and associated methods.
An efficient implementation to avoid multiple counting reduces the number of loops over the neighbours by a factor of 2 in two-dimensional simulations and by a factor of 6 in three-dimensional simulations.

\section{Numerical examples}\label{sec:examples}

The objective of this section is to illustrate the proposed theory through a set of numerical examples.
In addition to the examples that follow, we confirmed that the conditions~(\ref{eq:comp-10}) hold within numerical precision.
Four different studies are conducted.
In Section~\ref{subsec:examples-1}, a comparative study is carried out to investigate the influence of the horizon size and grid-spacing on the material response.
Then, the properties of the stiffness matrix are analysed for various parameters and geometries in Section~\ref{subsec:examples-2}.
This is followed by a two-part example in Section~\ref{subsec:examples-3} to study the Poisson effect in CPD for both two-dimensional and three-dimensional problems and also to compare CPD with CCM.
This comparison not only emphasises the similarities between CPD and CCM but it also provides a tangible case study to better understand the physical interpretation of the CPD material parameters and their role in describing common mechanical behaviour of materials.
Finally, in Section~\ref{subsec:examples-4}, a series of simulations at finite deformations are performed for both two-dimensional and three-dimensional domains.
These simulations demonstrate the influence of multi-neighbour interactions on the material response and illustrate the robustness of the framework and its consistent quadratic convergence even at very large deformations.

\subsection{Convergence and non-locality in CPD}\label{subsec:examples-1}

\begin{figure}[h]
    \centering
    \includegraphics[width=1.0\textwidth]{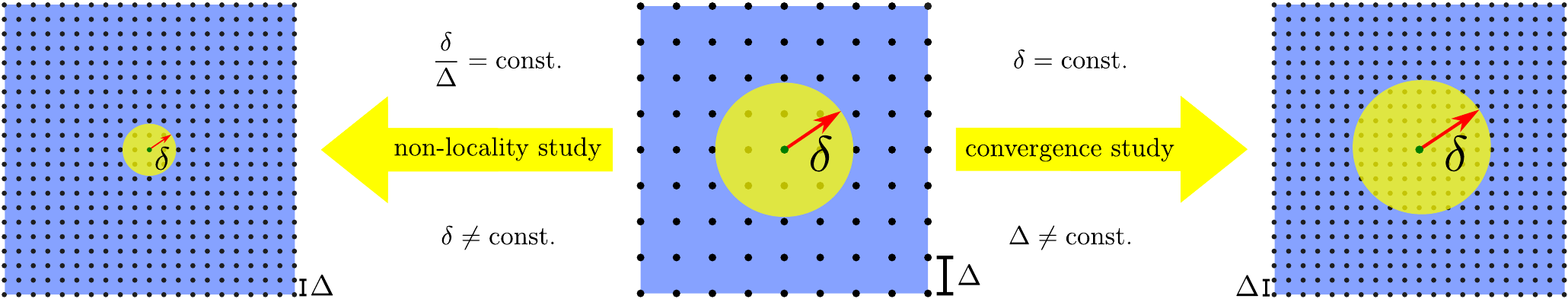}
    \caption{
    Schematic of a convergence study versus a non-locality study.
    For the convergence study, the horizon size $\delta$ remains constant while the grid-spacing is decreased.
    For the non-locality study, the ratio of the horizon-over-grid size $\delta/\Delta$ is fixed while the horizon is decreased to obtain a more local solution.
    }
    \label{fig:ex-1-00}
\end{figure}

The main goal of this section is to investigate the convergence behaviour and the inherent non-locality of the proposed framework.
This numerical example is carried out at small deformations in order to focus on the main objectives of this study.
A schematic of the investigation and a depiction of the horizon size $\delta$ and grid-spacing $\Delta$ is illustrated in Fig.~\ref{fig:ex-1-00}.
Furthermore, for all the computations in this section we utilize only one-neighbour interactions, for the sake of clarity.
We have carried out similar studies including, in addition, two-neighbour interactions and have observed similar trends.
The influence of two-neighbour interactions as well as three-neighbour interactions are studied separately in Section~\ref{subsec:examples-3} and \ref{subsec:examples-4}.

\begin{figure}[h]
    \centering
    \includegraphics[width=1.0\textwidth]{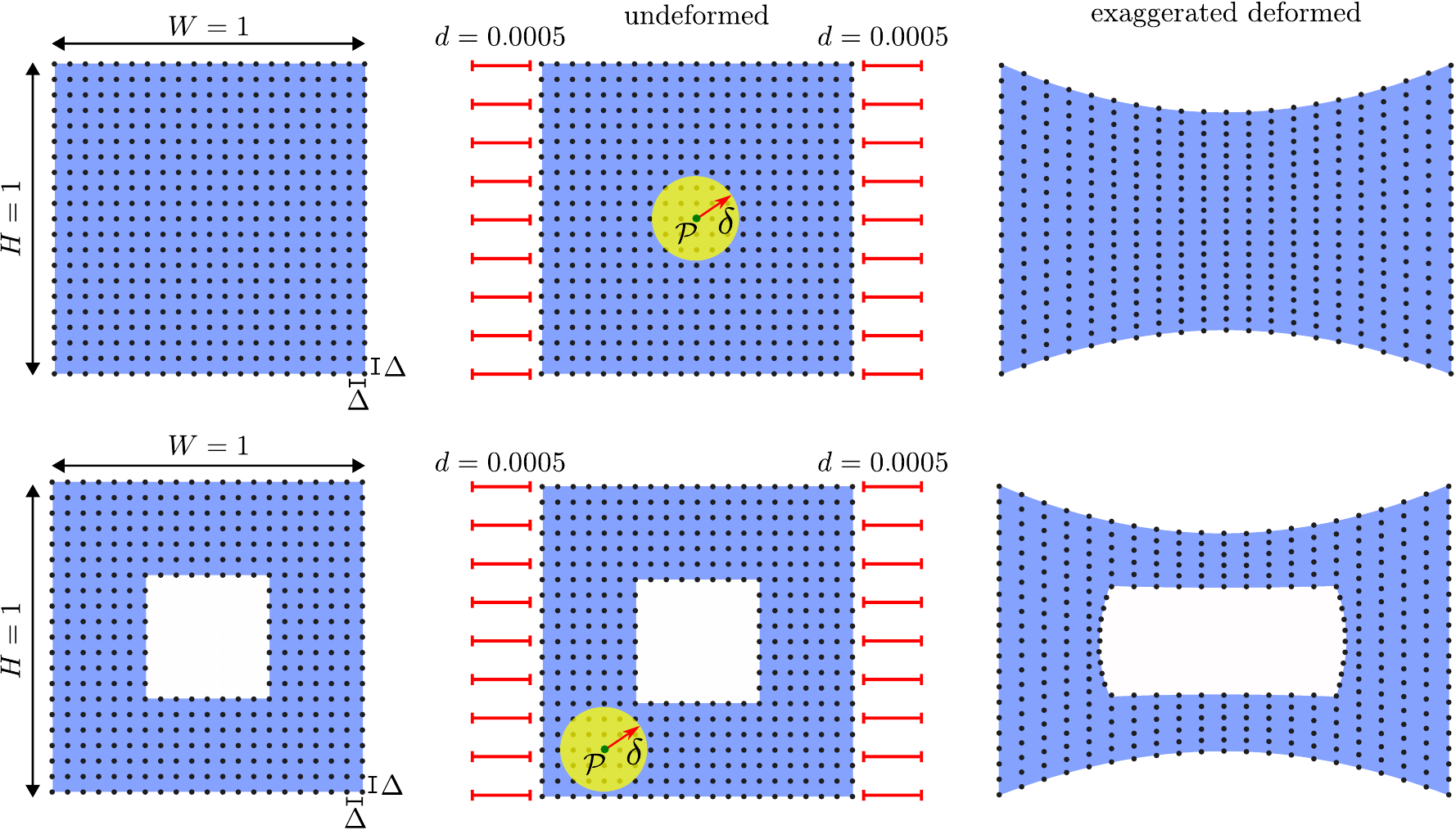}
    \caption{
    Schematic illustration of the specimens and the prescribed deformation.
    The first specimen is a full unit square and the second specimen is a unit square with a square hole at its centre.
    A lateral extension of $0.1\%$ is applied to the specimens.
    The exaggerated deformed shapes are depicted schematically on the right.
    }
    \label{fig:ex-1-0}
\end{figure}

The details of the two boundary value problems studied are shown in Fig.~\ref{fig:ex-1-0} with an \emph{exaggerated} deformation depicted.
For both examples, the grid points are uniformly distributed with horizontal and vertical spacing $\Delta$.
In the first row of Fig.~\ref{fig:ex-1-0}, a solid unit square is extended by $0.1\%$ in the horizontal direction.
In the second row, 40\% of the central region of the specimen is removed and the domain again extended by $0.1\%$ in the horizontal direction.
For each of the two domains, we carry out two separate studies.
For the first we fix the horizon size $\delta$ and decrease the grid-spacing $\Delta$ resulting in more neighbours within the horizon of each point.
This study demonstrates that, by increasing the number of grid points within the horizon, the solution of CPD converges, as expected.
In other words, for a given horizon size $\delta$, decreasing the grid-spacing $\Delta$ results in a more accurate solution.
In the second set of examples, the number of neighbours within the horizon remains constant but the horizon size itself varies.
More precisely, we change $\delta$ for a given $\delta/\Delta$ to study the non-locality associated with CPD.
We expect to observe decreased non-local effects with diminishing horizon size $\delta$.
In the limit of $\delta\to0$, we expect the solution of CPD to converge to the solution of CCM.
We include the solution from CCM, obtained using the finite element method, for comparison.

\begin{figure}[!h]
    \centering
    \includegraphics[height=0.9\textheight]{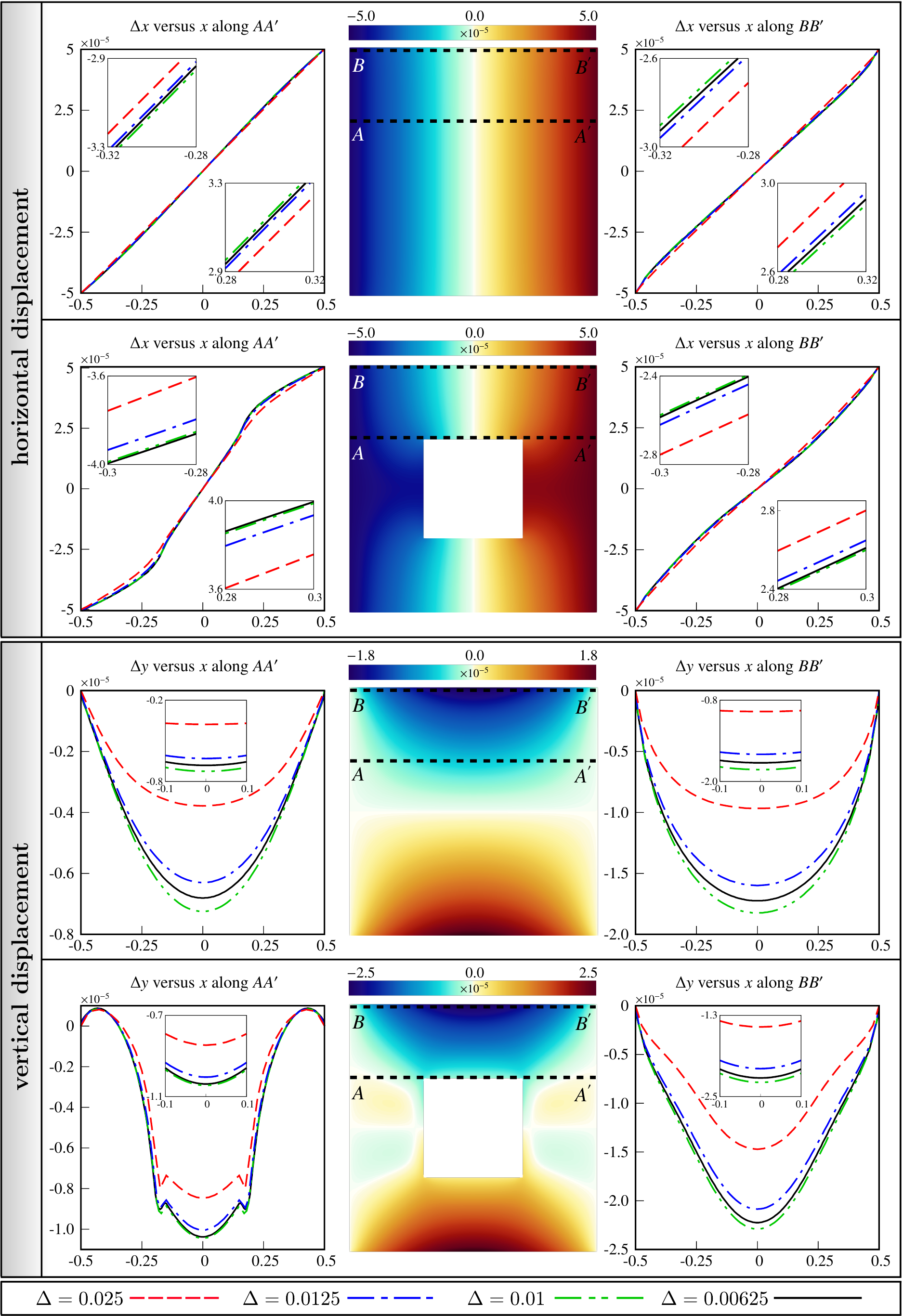}
    \caption{
    Illustration of the horizontal and vertical displacements throughout a full specimen and a specimen with a square hole under $0.1\%$ lateral extension.
    In this example, the horizon size $\delta$ is fixed and the grid-spacing $\Delta$ varies.
    }
    \label{fig:EXMP_1_1}
\end{figure}

\begin{figure}[!h]
    \centering
    \includegraphics[height=0.9\textheight]{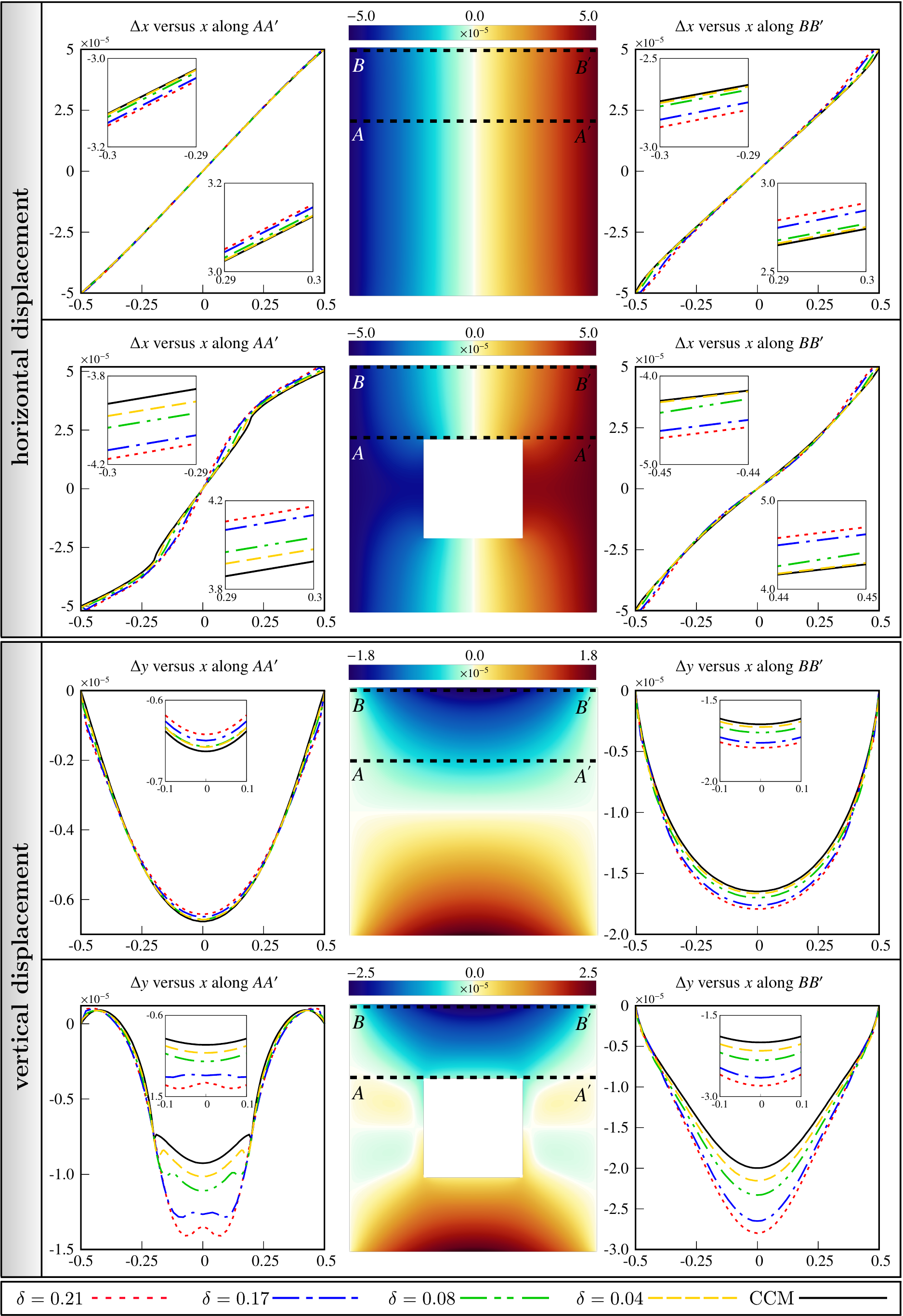}
    \caption{
    Illustration of the horizontal and vertical displacements throughout a full specimen and a specimen with a square hole under extension.
    In this example, the horizon size $\delta$ is varied together with the grid-spacing $\Delta$ while maintaining a fixed ratio of $\delta/\Delta = 8.5$.
    The solution corresponding to CCM included for the sake of comparison.
    }
    \label{fig:EXMP_1_2}
\end{figure}

Figure~\ref{fig:EXMP_1_1} shows the results for the convergence study.
The graphs and coloured distributions show the displacements in both horizontal and vertical directions due to the prescribed extension.
Four different values for the grid-spacing ($0.025$, $0.0125$, $0.01$ and $0.00625$) are considered and the horizon size is fixed at $\delta=0.05$.
The upper segment displays the horizontal displacement and the lower segment the vertical displacement.
The distribution of the displacement throughout the specimen is depicted at the centre of each row.
The distribution of the solution over the lines $AA'$ and $BB'$ is investigated; $AA'$ is located at the top of the hole and $BB'$ is located on the upper edge of the specimens.
The figures on the left and right side of each row show the displacement along these two lines.
For both domains, the computational results are in excellent agreement with our previously stated expectations.
In short, the results converge upon reduction in the grid-spacing $\Delta$ for a given horizon $\delta$.
Reducing the grid-spacing while fixing the horizon size increases the number of neighbours for each point leading to a more accurate solution, hence the convergence observed.
It should be emphasised that the vertical displacement occurs solely due to the Poisson effect.
The vertical displacement of the full specimen decreases from zero to an extremum in the middle and are symmetric about both $AA'$ and $BB'$.
In addition, the vertical displacement over the line $BB'$ is greater than over $AA'$, due to its increased distance from the centreline.
The vertical displacement behaviour of the specimen with a square hole is more complicated.
Over the line $BB'$, we observe a decrease from the edge to the middle of the domain, whereas the vertical displacement along $AA'$ increases at the beginning and then decreases to an extremum in the middle.
This trend is not an artifice of our CPD formulation and is also observed in CCM.
Moreover, the vertical displacements along both lines exceed those obtained for the full specimen, which again can be explained due to the more pronounced Poisson effect for a specimen with a hole at its centre.

\begin{figure}[!h]
    \centering
    \includegraphics[height=0.9\textheight]{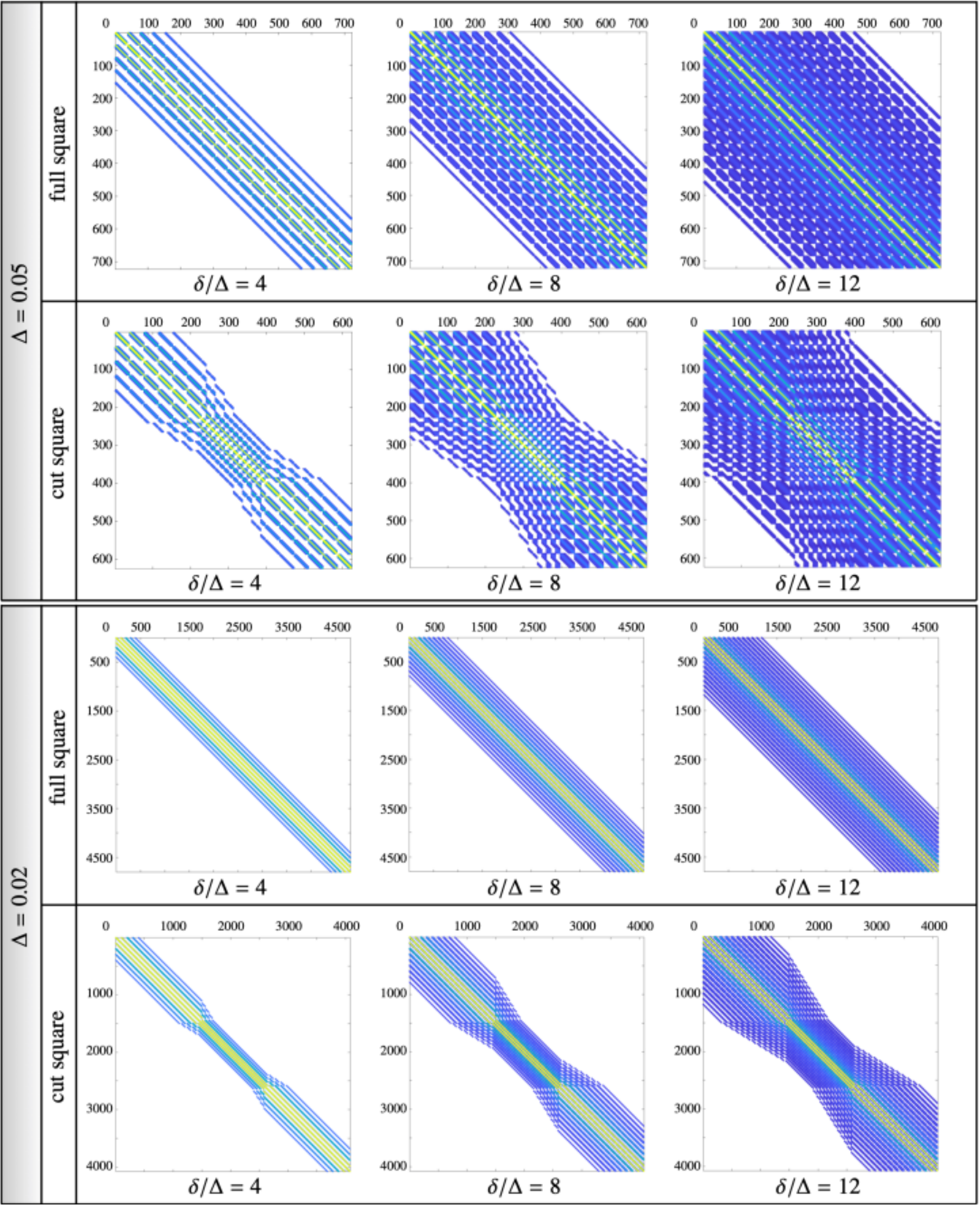}
    \caption{
    Depiction of the sparsity patterns of the stiffness matrix for both the full domain and the domain with a square hole for different grid-spacings $\Delta$, as well as different horizon-over-grid size $\delta/\Delta$ ratios.
    The stiffness matrix is clearly symmetric.
    The colours correspond to the absolute values of the components.
    }
    \label{fig:EXMP-0}
\end{figure}

In the second set of numerical studies, the horizon-over-grid size $\delta/\Delta$ is fixed and the horizon size $\delta$ varies to highlight the non-local nature of CPD.
Figure~\ref{fig:EXMP_1_2} shows both the vertical and horizontal displacements throughout the specimens.
To facilitate comparison, the problem is designed to have the same features as the first numerical study depicted in Fig.~\ref{fig:EXMP_1_1}.
Four different horizon sizes ($\delta=0.21$, $\delta=0.17$, $\delta=0.08$ and $\delta=0.04$) are considered and $\delta/\Delta=8.5$.
The solutions associated with CCM, obtained using the finite element method with a sufficiently fine mesh, are included for comparison.
The solutions obtained from CPD are denoted by a dashed line whereas a solid black line represents the solution corresponding to CCM.
Similar to the previous example, the  distribution of the displacements throughout the specimens is depicted at the centre of each row and the two figures to the left and right depict the displacements along the lines $AA'$ and $BB'$, respectively.
For all cases, decreasing the horizon size results in less deviation from the local solution associated with CCM thereby demonstrating the expected behaviour.
That is, we observe a decrease in non-local effects with diminishing horizon size $\delta$ and asymptotically, in the limit of $\delta\to0$, the solution of CPD converges to the solution of CCM.

\subsection{Properties of the stiffness matrix}\label{subsec:examples-2}

This section elaborates on properties of the stiffness matrix $\v{K}$ such as sparsity and symmetry.
Figure~\ref{fig:EXMP-0} shows the sparsity pattern of the stiffness matrix for the full specimen and the specimen with a square hole.
Two different grid-spacings $\Delta=0.02$ and $\Delta=0.05$ are examined and for each case several horizon-over-grid ratios $\delta/\Delta$ are considered.
The horizontal and vertical axis in each figure corresponds the columns and rows of the stiffness matrix, respectively.
As expected, the stiffness matrix is symmetric for all the cases.
The colours in Fig.~\ref{fig:EXMP-0} correspond to the absolute values of the components of the stiffness matrix.
For the specimen with the square hole, the stiffness matrix has a narrower bandwidth in the vicinity of the hole since fewer neighbours are present for each point in that region.
Increasing the ratio $\delta/\Delta$ results in a larger bandwidth and more non-zero values (decreased sparsity) in the stiffness matrix.
This can be explained as by increasing $\delta/\Delta$, more points contribute to each  degree of freedom.
For the case with $\Delta=0.02$, although the pattern seems visually narrower than the case with $\Delta=0.05$, the number of degrees of freedom are different.
There is a significant difference between the ranges for these two cases indicating that the stiffness matrix is considerably larger when $\Delta=0.02$, which is intuitive since a smaller grid-spacing $\Delta$ translates to more points, and hence more degrees of freedom.

 \subsection{Interplay between Poisson effect and the material parameters}\label{subsec:examples-3}

This section provides a detailed comparison between CPD and CCM with a focus on the Poisson effect and its relation to the material parameters associated with each theory.
The material parameters of CCM are the Lam{\'e} parameters $\lambda$ and $\mu$ while for CPD they are $C_{1}$, $C_{2}$ and $C_{3}$.
The parameters $C_1$, $C_2$ and $C_3$ correspond to one-neighbour, two-neighbour and three-neighbour interactions, respectively.
Conceptually, $C_1$, $C_2$ and $C_3$ can be interpreted as resistance against the change of length, area and volume, respectively.
Both two- and three-dimensional analyses are carried out and the significance of $C_2$ and $C_3$ for two-dimensional and three-dimensional problems explained.
To perform this study, the specimen is subject to an extension at small deformations and the effective Poisson ratio is calculated.
It should be emphasised that the effective Poisson ratio for both CPD and CCM is treated as a geometrical feature computed via numerical simulation.
That is, the Poisson ratio is calculated by dividing the lateral contraction by the extension at the centre of the domain.
Note that for the CPD analysis, the specimen is discretised into grid (collocation) points, whereas for the CCM analysis the specimen is discretised by finite elements.
The finite elements are bilinear quadrilaterals and trilinear hexahedrals for the two-dimensional and three-dimensional computations, respectively.

\begin{figure}[!h]
    \centering
    \includegraphics[width=0.9\textwidth]{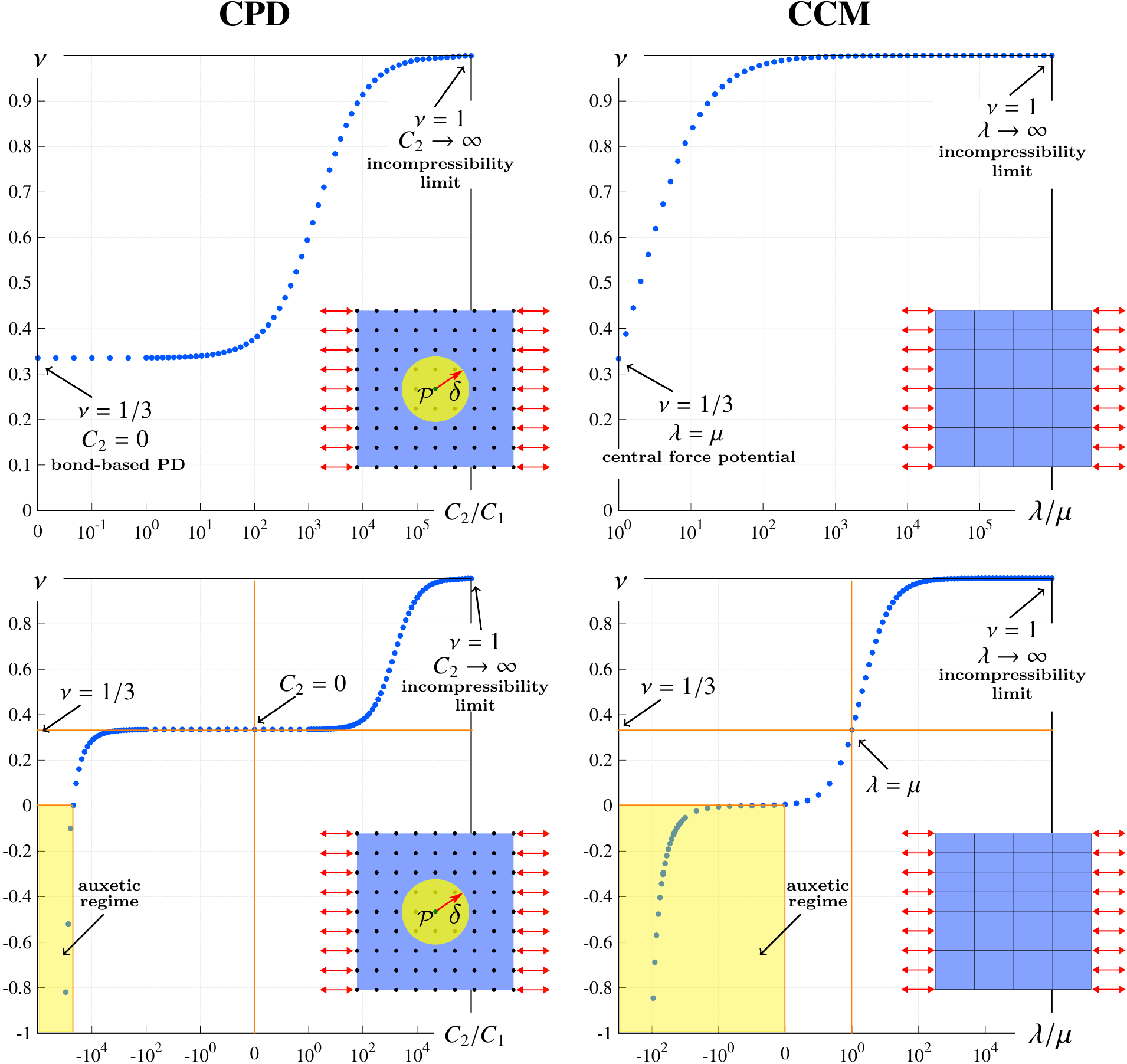}
    \caption{
    Variation of the Poisson ratio versus material properties in continuum-kinematics-inspired peridynamics and classical continuum mechanics for a two-dimensional case.
    The left figures correspond to continuum-kinematics-inspired peridynamics (CPD) and the right figures correspond to classical continuum mechanics (CCM).
    }
    \label{fig:EXMP_3_1}
\end{figure}

\begin{figure}[!h]
    \centering
    \includegraphics[width=0.9\textwidth]{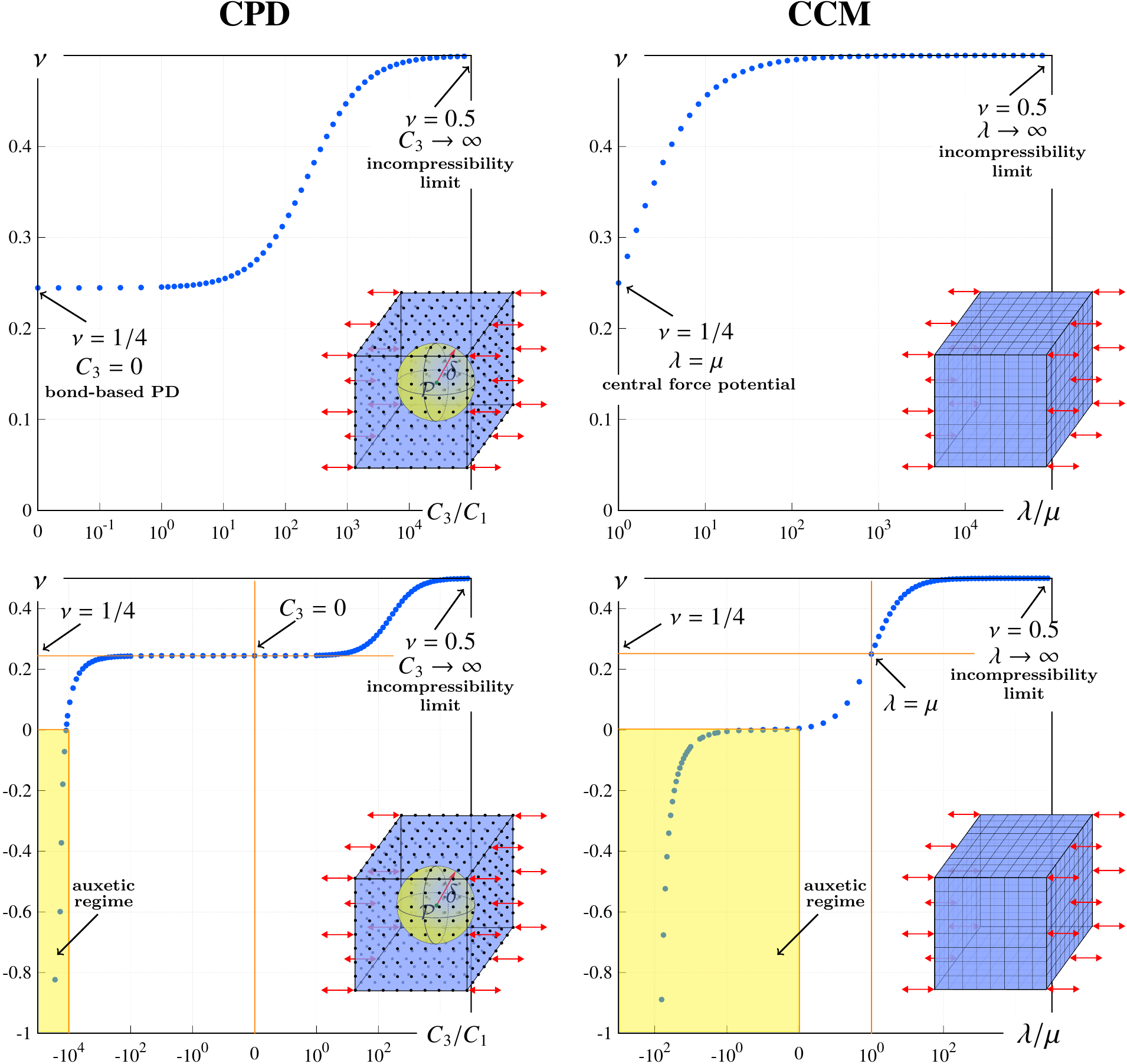}
    \caption{
    Variation of the Poisson ratio versus material properties in continuum-kinematics-inspired peridynamics and classical continuum mechanics for a three-dimensional case.
    The left figures correspond to continuum-kinematics-inspired peridynamics (CPD) and the right figures correspond to classical continuum mechanics (CCM).
    }
    \label{fig:EXMP_3_3}
\end{figure}

\begin{rmk}
It is important to recall that both CPD and CCM require three constants for isotropic elasticity at finite deformations.
However, the linearisation process at small strains reduces the number of independent parameters from three to two in CCM.
The CPD formalism accounts directly for finite deformations and is not a linearised theory, and hence the three constants are present.
It would be feasible to establish a linear CPD theory in which only two independent parameters contribute to the material behaviour. \qed
\end{rmk}

\begin{figure}[b!]
    \centering
    \includegraphics[width=1.0\textwidth]{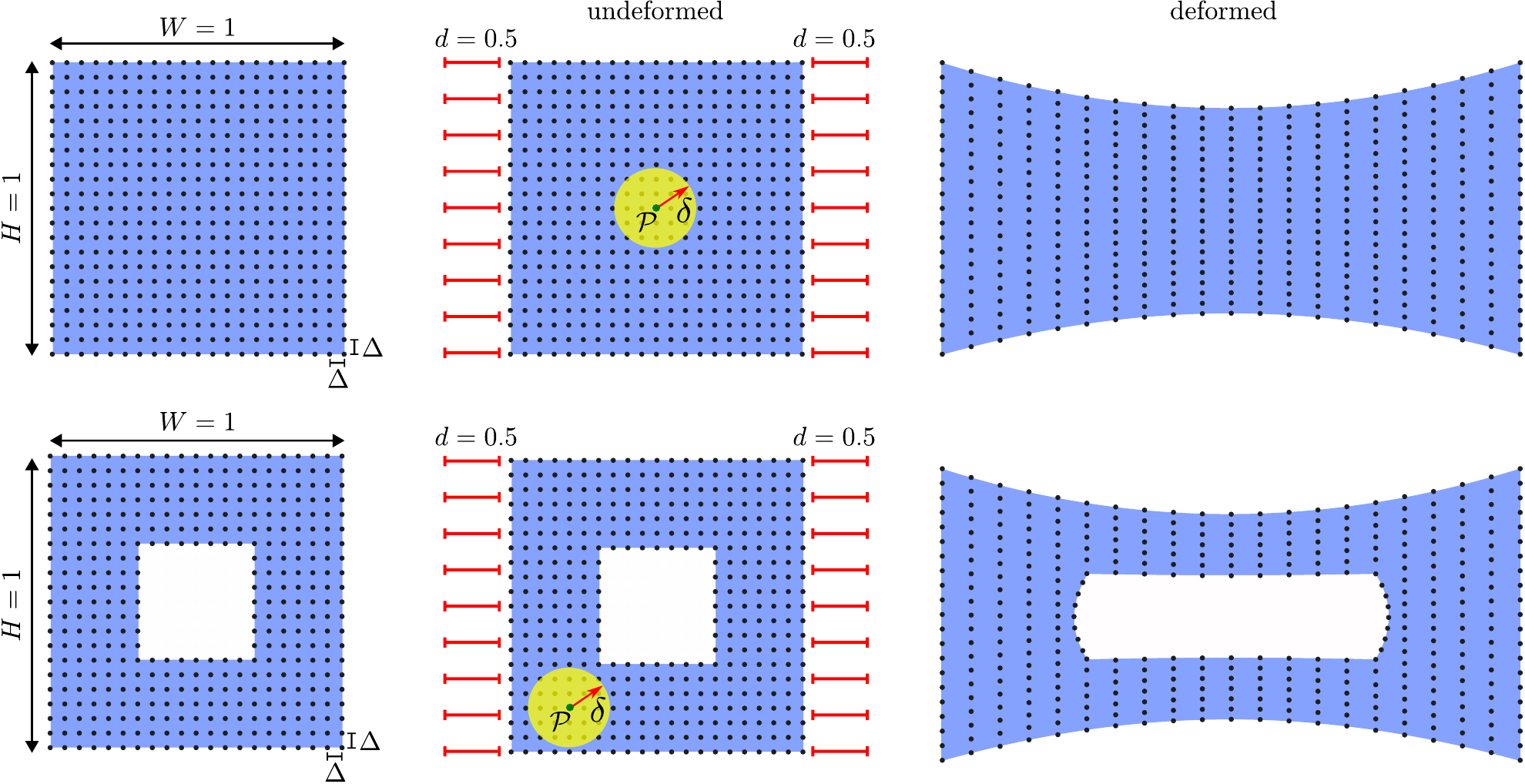}
    \caption{
    Schematic illustration of the two finite deformation example problems.
    The first specimen is a unit square and the second  a unit square with a square hole at its centre.
    Both specimens are subject to an extension of $100\%$.
    }
    \label{fig:EX_4_0}
\end{figure}

Figure~\ref{fig:EXMP_3_1} illustrates the Poisson ratio versus the material parameters for a two-dimensional problem.
The two top figures are a magnified portion of the bottom two and are provided for the sake of clarity.
The bottom figures sweep a broader range of the material parameters so as to cover the whole range of permissible Poisson ratios including the auxetic regime.
For the CPD analysis, the horizontal axis represents the ratio of the two-neighbour elastic coefficient to the one-neighbour elastic coefficient $C_{2}/C_{1}$ whereas for the CCM analysis it corresponds to the ratio of the first to the second Lam{\'e} parameter $\lambda/\mu$.
The parameters $C_{1}$ and $\mu$ are set to $1$ and we vary $C_{2}$ and $\lambda$ to generate the desired range for the ratios.
Setting $C_{2}=0$ implies that two-neighbour interactions do not contribute.
Thus only the one-neighbour interactions of CPD are active corresponding to bond-based peridynamics.
It is observed that a Poisson ratio of $1/3$ is obtained in CPD when $C_{2}=0$.
This is well-known for bond-based PD~\cite{Madenci2014}.
A Poisson ratio of $1/3$ is obtained in CCM when $\lambda=\mu$ as expected according to the relation $\nu=\lambda/[\lambda+2\mu]$ for two-dimensional local (linear) elasticity.
Increasing $C_{2}$ or $\lambda$ results in a higher resistance to a change of area, and hence we approach the incompressible limit.
In the limits of $C_{2} \to \infty$ and $\lambda \to \infty$, the two-dimensional incompressibility limit $\nu=1$ is obtained.
It is observed that although there is a one-to-one relation between the graphs on the right and the ones on the left, for a given Poisson ratio, $C_{2}/C_{1}$ does not coincide exactly with the same value for $\lambda/\mu$.
However, the Poisson ratio corresponding to $C_{2}/C_{1}=0$ is identical to that associated with $\lambda/\mu=0$.

Figure~\ref{fig:EXMP_3_3} shows the Poisson ratio versus the material parameters for a three-dimensional domain.
The horizontal axis for CPD, in contrast to the two-dimensional problem presented previously, corresponds to the ratio of the three-neighbour to the one-neighbour elastic coefficient $C_{3}/C_{1}$.
The horizontal axis for CCM, similar to the two-dimensional problem, is the ratio $\lambda/\mu$.
Similar to the previous case, the parameters $C_{1}$ and $\mu$ are set to $1$ and $C_{3}$ and $\lambda$ are calculated according to the desired range for the ratios on the horizontal axis.
Note that in order to examine the effect of three-neighbour interactions, the two-neighbour elastic coefficient $C_{2}$ is set to zero here.
In CPD, the Poisson ratio $1/4$ is recovered when $C_{3}=0$; this corresponds exactly to the bond-based PD, as expected.
In CCM, the same Poisson ratio is obtained when $\lambda=\mu$ which agrees directly with the classical relation $\nu=\lambda/[2\lambda+2\mu]$.
Similar to the two-dimensional study, in the limits of $C_{3} \to \infty$ and $\lambda \to \infty$, the incompressibility limit is obtained.
The incompressibility limit now corresponds to $\nu=0.5$ due to the three-dimensional nature of the problem.
As demonstrated in these two examples, our methodology can not only exactly recover bond-based peridynamics but it is also capable of covering the whole range of possible Poisson ratios including the auxetic regime.

\begin{figure}[b!]
    \centering
    \includegraphics[width=1.0\textwidth]{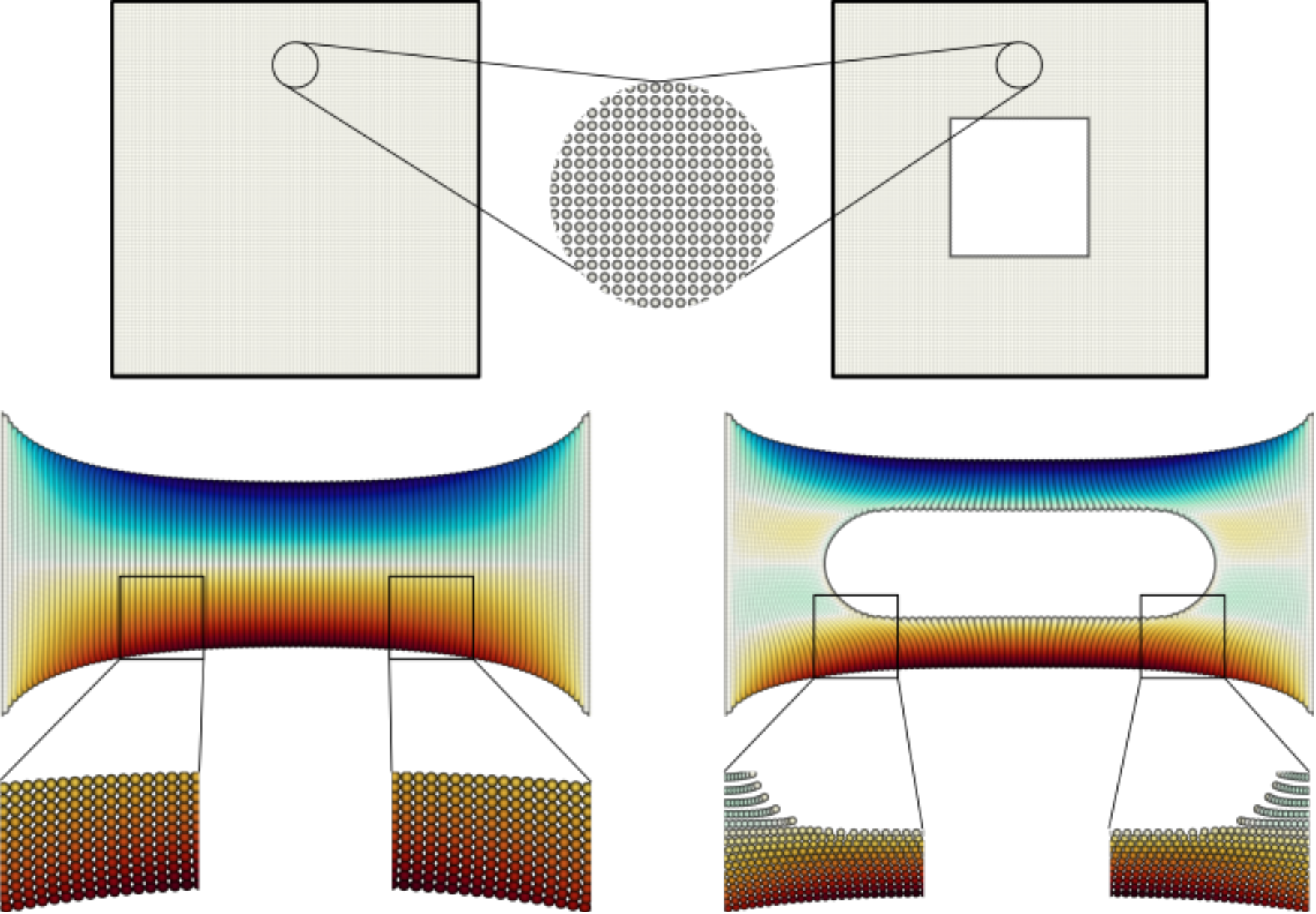}
    \caption{
    Illustration of the original specimens and their deformed shapes for the two dimensional example at finite deformations.
    Zoom boxes are included to elucidate the distribution of the points throughout the specimens.
   }
    \label{fig:EX_4_1}
\end{figure}

\begin{figure}[b!]
    \centering
    \includegraphics[width=0.92\textwidth]{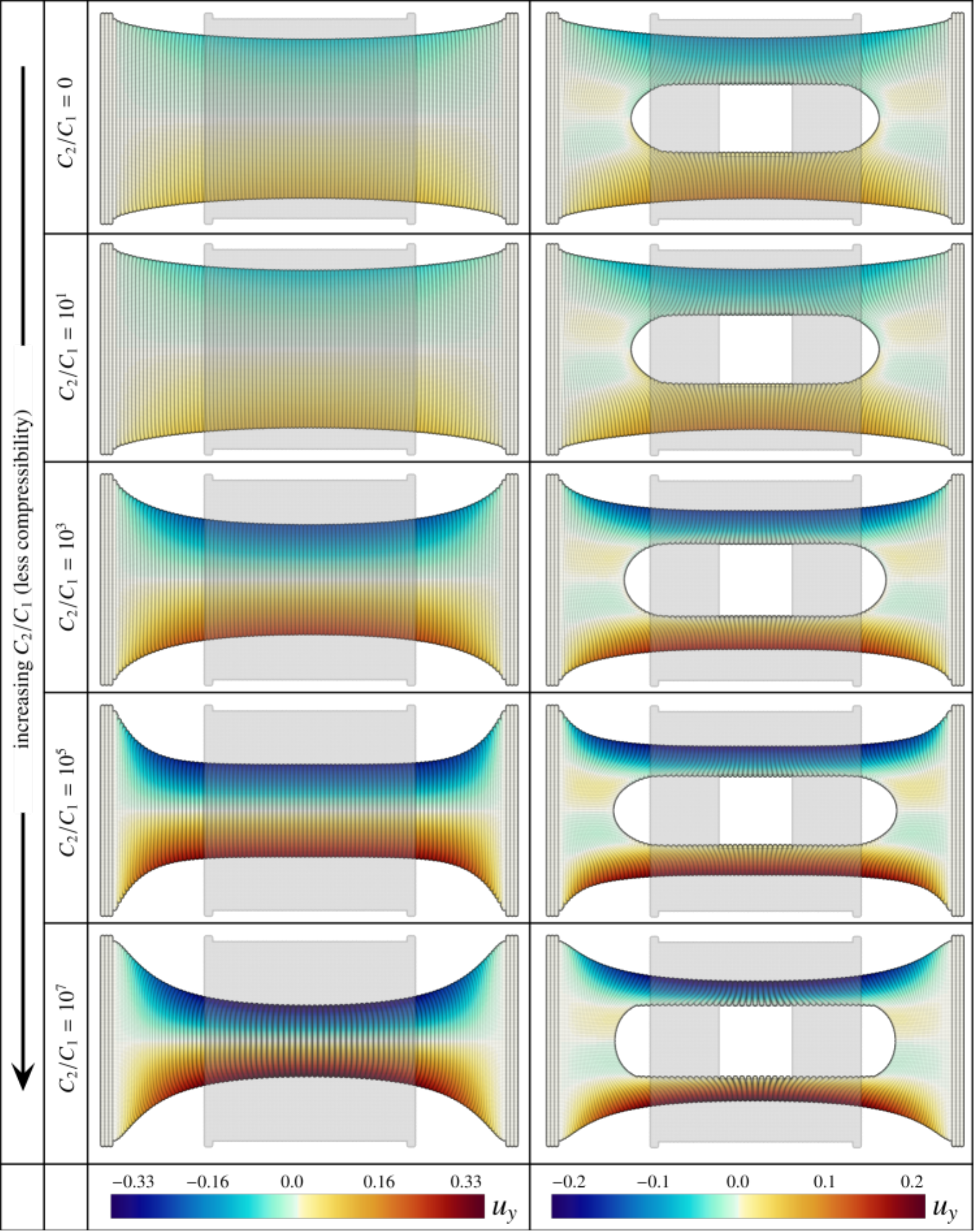}
    \caption{
    Large deformations of a unit square without and with a square hole at its centre for different $C_2/C_1$ ratio associated with different levels of incompressibility.
    }
    \label{fig:EX_4_2}
\end{figure}

\subsection{CPD at finite deformations}\label{subsec:examples-4}

As shown in the previous section, CPD is capable of capturing any Poisson ratio by incorporating two-neighbour and three-neighbour interactions.
In this section, we carry out a series of computational studies at finite deformations to compare the influence of multi-neighbour interactions on the material response for both two-dimensional and three-dimensional frameworks.
Furthermore, we demonstrate the robustness of the proposed framework.
For the two-dimensional analysis, two different specimens are considered with the geometry and loading conditions illustrated in Fig.~\ref{fig:EX_4_0}.
Both specimens are subject to $100\%$ extension in the horizontal direction and are free in the lateral direction.
We consider a unit square without and with a square hole at its centre as shown in the first and second rows of Fig.~\ref{fig:EX_4_0}, respectively.
The hole is in the shape of a square with the sides of length $0.4$.
For the three-dimensional analysis that follows, we mimic similar conditions and dimensions.
First we detail the numerical simulations for a two-dimensional domain.

The discretised specimens used in the analysis are depicted in Fig.~\ref{fig:EX_4_1} together with their deformed shapes obtained from computational simulations.
The shading corresponds to the vertical displacements and zoom boxes are provided for further clarity.
Figure~\ref{fig:EX_4_2} illustrates the deformed discrete domain for both specimens and for various values of $C_{2}/C_{1}$.
The transparent shapes depicts the undeformed configuration.
Prescribing displacement-type boundary conditions in CPD is similar to PD.
That is, we prescribe the displacements for a few layers of points on the boundary as shown.
The number of the layers over which the boundary conditions are imposed depends on the horizon size.
In the first row of Fig.~\ref{fig:EX_4_2}, $C_{2}=0$ and thus, only one-neighbour interactions are active.
The remaining rows involve both one-neighbour and two-neighbour interactions where $C_{2}/C_{1}$ provides a measure of incompressibility and hence, given the loading and geometry, larger values of $C_{2}/C_{1}$ lead to increased lateral contraction.

\begin{figure}[b!]
    \centering
    \includegraphics[width=1.0\textwidth]{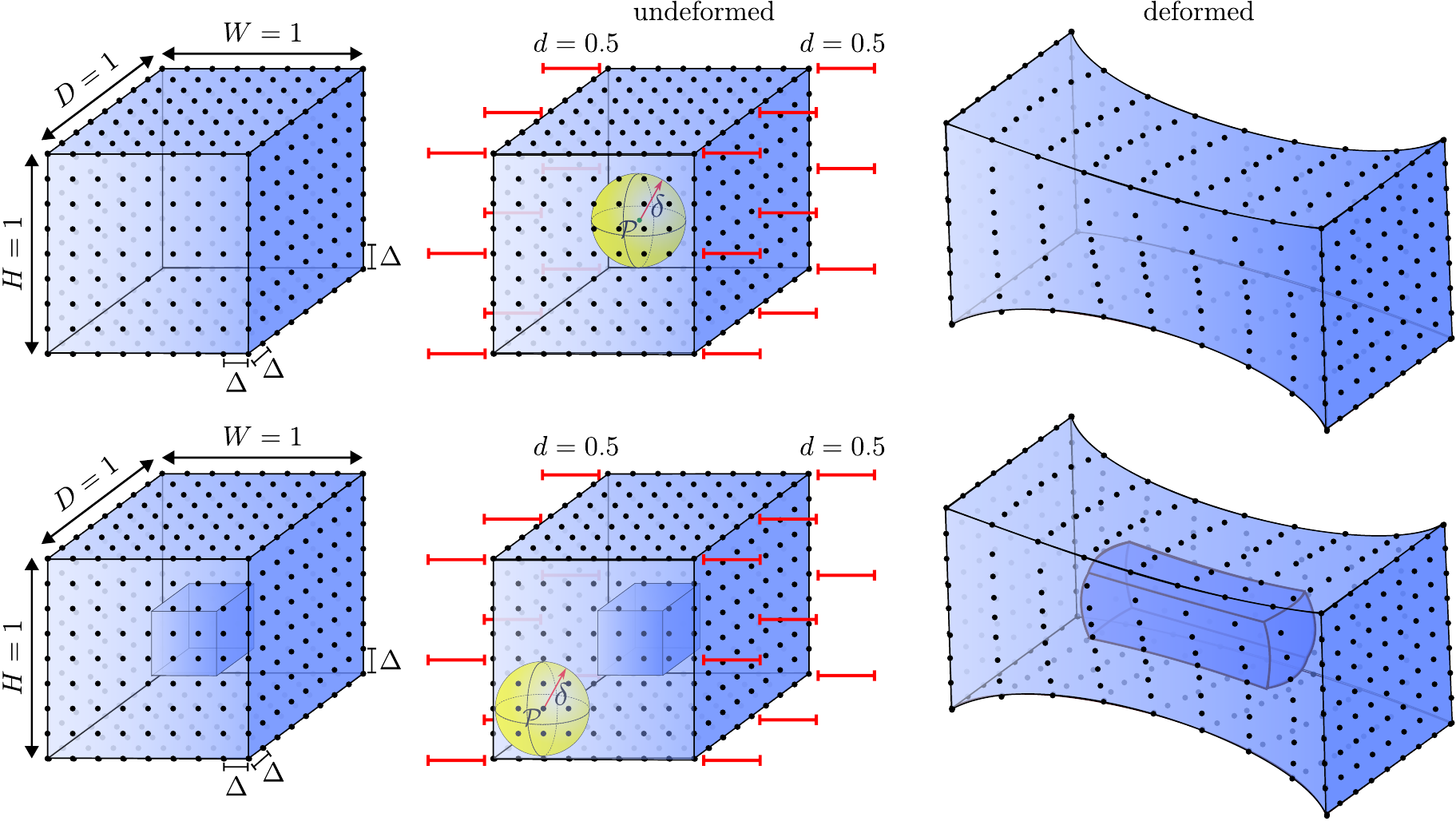}
    \caption{
    Schematic illustration of the two study cases undergone and the applied deformation type.
    The first specimen is a full unit cube and the second specimen is a unit cube with a cubic hole at its centre.
    Both specimens are subject to $100\%$ extension.
    }
    \label{fig:EX_4_00}
\end{figure}

Figure~\ref{fig:EX_4_00} presents a schematic illustration of the specimens and the prescribed deformations for the three-dimensional analysis.
Similar to the two-dimensional analysis, both specimens are subject to $100\%$ extension in the horizontal direction and are free in both lateral directions.
We consider a unit square without and with a cubic hole at its centre as shown in the first and second rows of Fig.~\ref{fig:EX_4_00}, respectively.
The hole is also in the shape of a cube with sides of dimension $0.4$.

\begin{figure}[h!]
    \centering
    \includegraphics[height=0.9\textheight]{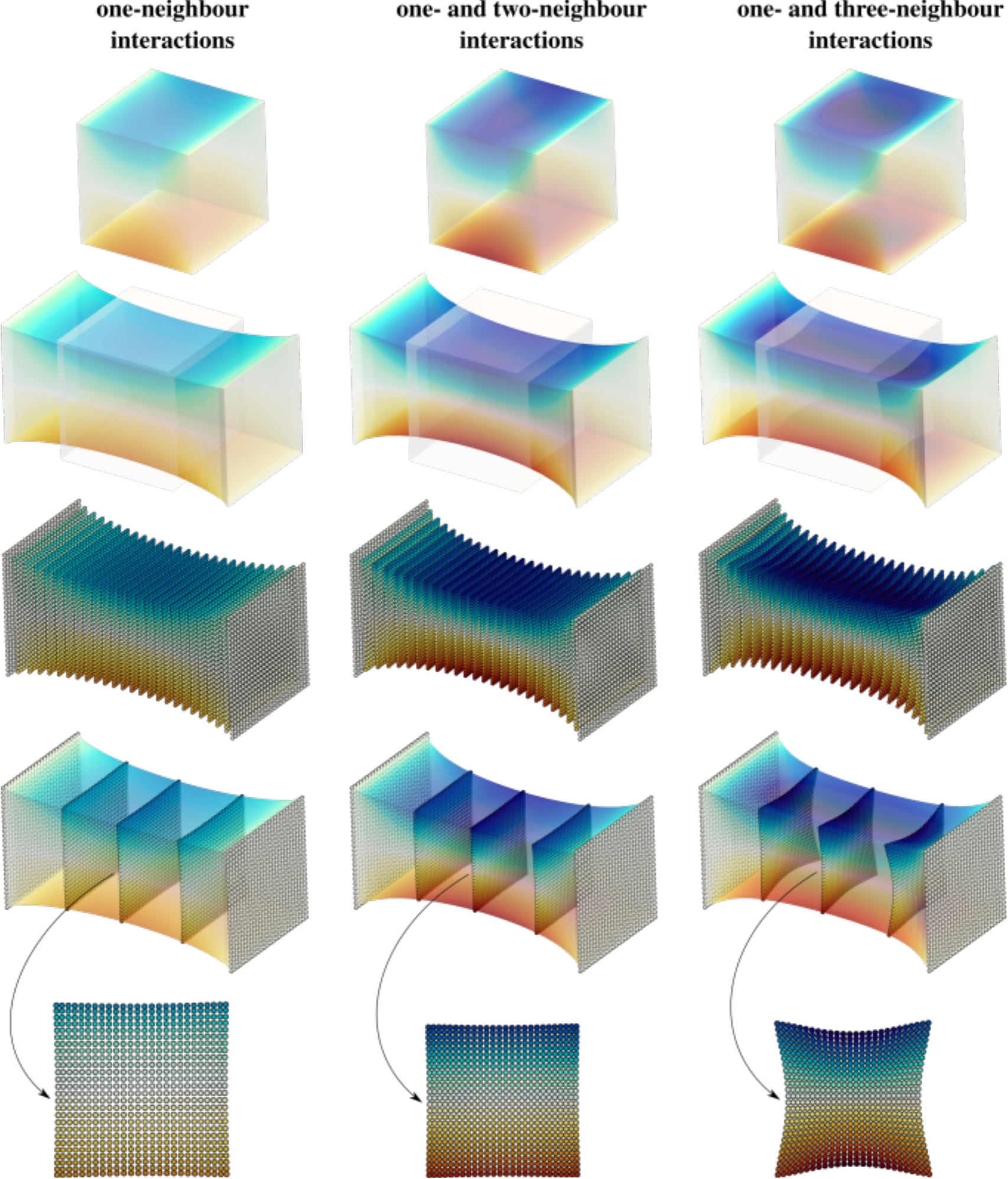}
    \caption{
    Large deformation of a cube for different types of interactions.
    }
    \label{fig:EX_4_3}
\end{figure}

\begin{figure}[h!]
    \centering
    \includegraphics[height=0.9\textheight]{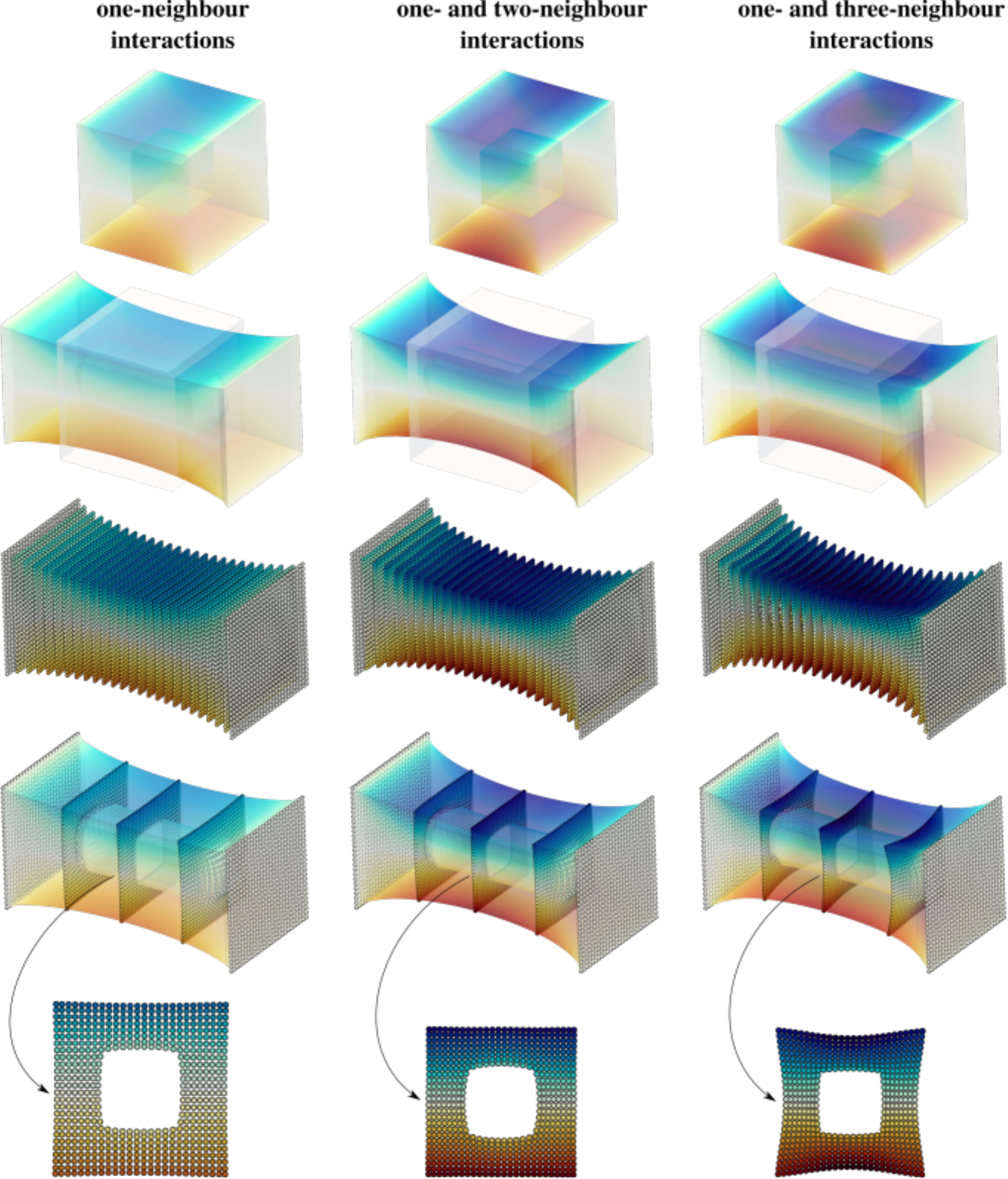}
    \caption{
    Large deformation of a cube with a cubic hole at its centre for different types of interactions.
    }
    \label{fig:EX_4_4}
\end{figure}

Figures~\ref{fig:EX_4_3} and~ \ref{fig:EX_4_4} show the deformed shapes as well as the distribution of the vertical displacement on both domains.
Three different interaction types are examined.
The first column is associated with one-neighbour interactions.
The results in the first column provide an excellent reference against which to compare the results in the second and third columns.
The second column takes one-neighbour and two-neighbour interactions into account whereas the third column accounts for one-neighbour and three-neighbour interactions.
These simulations are devised to distinguish between two-neighbour interactions and three-neighbour interactions.
We emphasise that it is not possible to have only two-neighbour interactions or only three-neighbour interactions.
This important observation will be contextualised later via a geometrical example.
In the first row, the reference configuration is illustrated where the colours correspond to the vertical displacement.
The deformed shapes are illustrated in the subsequent rows.
On the second row, both deformed and undeformed configurations are shown.
The smooth colour pattern is produced by interpolating the displacements between the points.
Hence, the third row is more representative of what we obtain from the simulations as no interpolation between points is assumed.
In the last row the deformed set of points on a vertical plane is extracted to highlight the difference between the three interaction types.
Since only one-neighbour interactions are considered in the 1st column, the least contraction is obtained which leads to increased compressibility when compared to the second and third columns.
It is important to note that although both two-neighbour interactions and three-neighbour interactions result in decreased compressibility, the solutions are different indicating that they indeed correspond to a different deformation.

\begin{figure}[h!]
    \centering
    \includegraphics[width=0.8\textwidth]{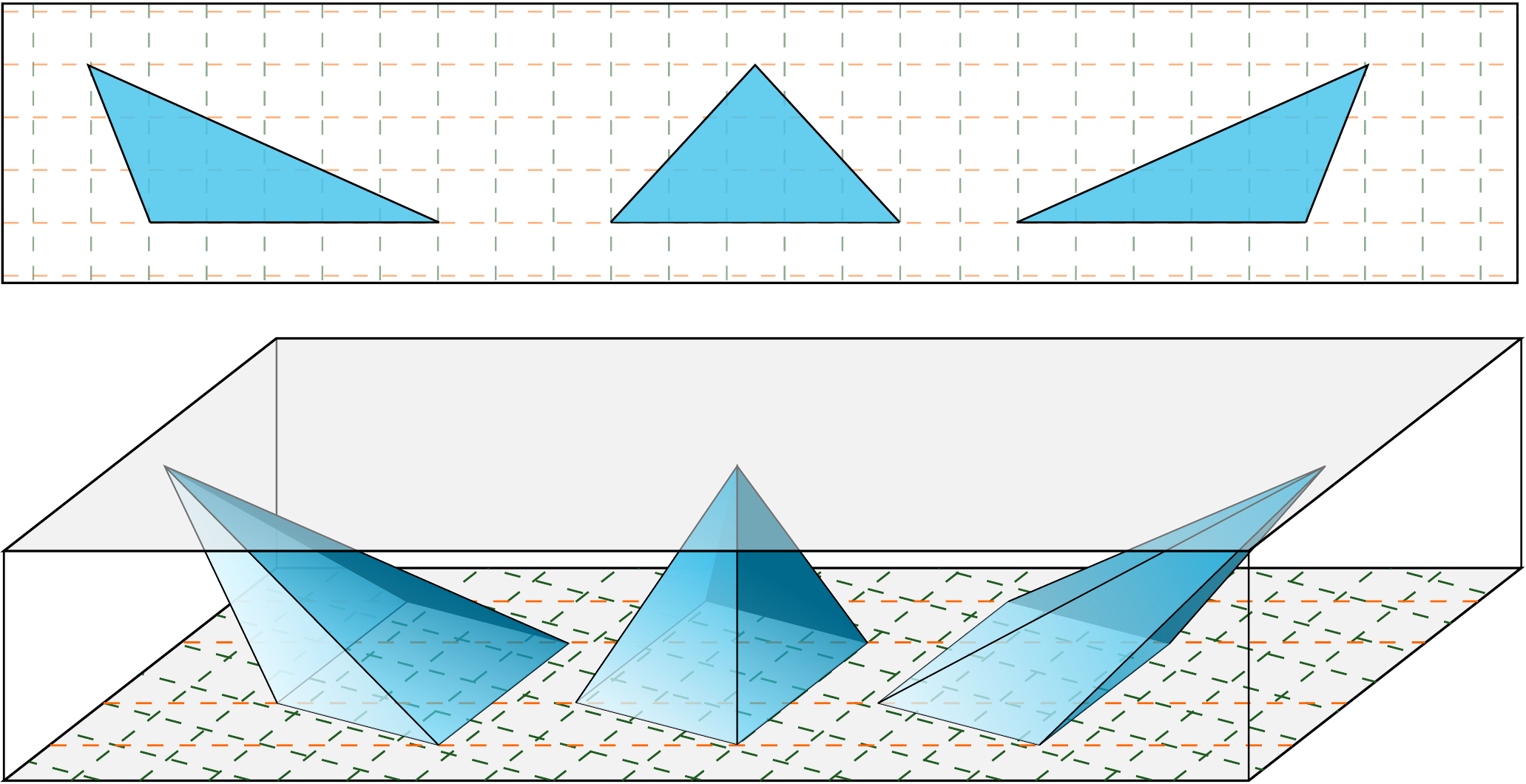}
    \caption{
    A toy example to explain the necessity of accounting for one-neighbour interaction, in contrast to two-neighbour and three-neighbour interactions.
    }
    \label{fig:EX_4_5}
\end{figure}

As mentioned, one-neighbour interactions are necessary for the stability of the computation.
In addition to the required one-neighbour interactions, two- and three-neighbour interactions can be included.
However, the opposite cannot be done.
More precisely, we cannot run simulations in which only two-neighbour or three-neighbour interactions are active and one-neighbour interactions are absent.
That is, while two-neighbour and three-neighbour interactions are important to capture the Poisson effect correctly, one-neighbour interaction are necessary to avoid instabilities.
The requirement for one-neighbour interactions is a physical one.
Consider Fig.~\ref{fig:EX_4_5}.
In the upper part of the image three triangles are shown with equal area.
In the lower half three pyramids are depicted with identical volumes.
Although the areas of the various triangles and the volumes of the various pyramids are the same, the length of their edges differ among them.
Hence in the absence of one-neighbour interactions one could have multiple configurations with identical energies that would in turn cause instabilities.
However, it is not possible to change a configuration without changing the distances between the pairs of points.
Thus one-neighbour interactions guarantee that different configurations correspond to different energies, which does not hold for only two-neighbour and three-neighbour interactions.

Finally, to illustrate the robustness of the proposed framework and its ability to simulate large deformations, we study the convergence behaviour.
Table~\ref{tab:convergence} gathers the convergence of the normalized residual for the three-dimensional simulations at large deformations shown in Fig.~\ref{fig:EX_4_3} for various interactions at different increments.
The domain in this study is subject to $100\%$ extension in $25$ increments and the ${\ell}^2$-norm of the residual $|\v{R}|$ recorded every fifth increment.
Despite the large deformation, consistent quadratic convergence is obtained consistently at every increment.

\begin{table}[h!]
\setstretch{1.0}
   \caption{Quadratic convergence in CPD. The numbers indicate the normalised norm of the residual $\v{R}$ at different increments for various types of interactions.}
   \label{tab:convergence}
   \small 
   \centering 
   \begin{tabular}{|p{0.06\textwidth}||p{0.12\textwidth}|p{0.12\textwidth}|p{0.12\textwidth}|p{0.12\textwidth}|p{0.12\textwidth}|p{0.12\textwidth}|p{0.12\textwidth}|}
   \toprule[\heavyrulewidth]\toprule[\heavyrulewidth]
     & Increment 1 & Increment 5 & Increment 10 & Increment 15 & Increment 20 & Increment 25\\
   \midrule
   \begin{tabular}{@{}c@{}} \multirow{5}{*}{\rotatebox{90}{one-}} \end{tabular}
   \begin{tabular}{@{}c@{}} \multirow{5}{*}{\rotatebox{90}{neighbour}} \end{tabular}
   \multirow{5}{*}{\rotatebox{90}{interactions}}
	& 1          & \cellcolor{gray}1          & 1          & \cellcolor{gray}1          & 1          & \cellcolor{gray}1          \\
	& 8.97\,e-02 & \cellcolor{gray}7.43\,e-02 & 5.91\,e-02 & \cellcolor{gray}4.72\,e-02 & 3.80\,e-02 & \cellcolor{gray}3.08\,e-02 \\
	& 1.14\,e-03 & \cellcolor{gray}6.08\,e-04 & 2.88\,e-04 & \cellcolor{gray}1.42\,e-04 & 7.38\,e-05 & \cellcolor{gray}3.98\,e-05 \\
	& 2.41\,e-07 & \cellcolor{gray}5.36\,e-08 & 9.46\,e-09 & \cellcolor{gray}1.92\,e-09 & 4.51\,e-10 & \cellcolor{gray}1.20\,e-10 \\
	& 1.23\,e-14 & \cellcolor{gray}6.36\,e-15 & 6.83\,e-15 & \cellcolor{gray}7.23\,e-15 & 7.69\,e-15 & \cellcolor{gray}8.13\,e-15 \\
   \bottomrule[\heavyrulewidth]
   \end{tabular}\\
   \begin{tabular}{|p{0.06\textwidth}||p{0.12\textwidth}|p{0.12\textwidth}|p{0.12\textwidth}|p{0.12\textwidth}|p{0.12\textwidth}|p{0.12\textwidth}|p{0.12\textwidth}|}
   \toprule[\heavyrulewidth]
   \begin{tabular}{@{}c@{}} \multirow{5}{*}{\rotatebox{90}{one- and two-}} \end{tabular}
   \begin{tabular}{@{}c@{}} \multirow{5}{*}{\rotatebox{90}{neighbour}} \end{tabular}
   \multirow{5}{*}{\rotatebox{90}{interactions}}
	& 1          & \cellcolor{gray}1          & 1          & \cellcolor{gray}1          & 1          & \cellcolor{gray}1          \\
	& 8.23\,e-02 & \cellcolor{gray}7.12\,e-02 & 6.02\,e-02 & \cellcolor{gray}5.16\,e-02 & 4.57\,e-02 & \cellcolor{gray}4.23\,e-02 \\
	& 9.73\,e-04 & \cellcolor{gray}6.48\,e-04 & 4.17\,e-04 & \cellcolor{gray}3.00\,e-04 & 3.45\,e-04 & \cellcolor{gray}1.34\,e-03 \\
	& 1.94\,e-07 & \cellcolor{gray}8.35\,e-08 & 3.43\,e-08 & \cellcolor{gray}2.07\,e-08 & 8.34\,e-08 & \cellcolor{gray}2.83\,e-06 \\
	& 1.17\,e-14 & \cellcolor{gray}6.91\,e-15 & 7.27\,e-15 & \cellcolor{gray}7.67\,e-15 & 1.19\,e-14 & \cellcolor{gray}1.56\,e-11 \\
   \bottomrule[\heavyrulewidth]
   \end{tabular}\\
   \begin{tabular}{|p{0.06\textwidth}||p{0.12\textwidth}|p{0.12\textwidth}|p{0.12\textwidth}|p{0.12\textwidth}|p{0.12\textwidth}|p{0.12\textwidth}|p{0.12\textwidth}|}
   \toprule[\heavyrulewidth]
   \begin{tabular}{@{}c@{}} \multirow{5}{*}{\rotatebox{90}{one- and three-}} \end{tabular}
   \begin{tabular}{@{}c@{}} \multirow{5}{*}{\rotatebox{90}{neighbour}} \end{tabular}
   \multirow{5}{*}{\rotatebox{90}{interactions}}
	& 1          & \cellcolor{gray} 1           & 1           & \cellcolor{gray} 1           & 1          & \cellcolor{gray} 1            \\
	& 8.97\,e-02 & \cellcolor{gray} 7.43\,e-02  & 5.91\,e-02  & \cellcolor{gray} 4.72\,e-02  & 3.80\,e-02 & \cellcolor{gray} 3.08\,e-02   \\
	& 1.14\,e-03 & \cellcolor{gray} 6.07\,e-04  & 2.88\,e-04  & \cellcolor{gray} 1.43\,e-04  & 7.51\,e-05 & \cellcolor{gray} 4.09\,e-05   \\
	& 2.40\,e-07 & \cellcolor{gray} 5.35\,e-08  & 9.47\,e-09  & \cellcolor{gray} 1.97\,e-09  & 4.82\,e-10 & \cellcolor{gray} 1.31\,e-10   \\
	& 1.24\,e-14 & \cellcolor{gray} 6.35\,e-15  & 6.83\,e-15  & \cellcolor{gray} 7.13\,e-15  & 7.85\,e-15 & \cellcolor{gray} 8.07\,e-15   \\
   \bottomrule[\heavyrulewidth]
   \end{tabular}\\
\end{table}

\FloatBarrier

\section{Conclusion}\label{sec:conclusion}

Continuum-kinematics-inspired peridynamics (CPD) was recently proposed by Javili et~al.~\citep{Javili2019} as a \emph{geometrically exact} alternative to peridynamics (PD)  to formulate non-local continuum mechanics at finite deformations.
Computational aspects of the CPD formulation have been presented for the first time.
The potential of the method has been made clear via a series of numerical examples.

A key feature of the proposed methodology is that it is fully implicit.
Furthermore, the tangent stiffness is computed directly and not via numerical differentiation schemes, unlike the commonly accepted strategy in classical state-based peridynamics.
The numerical implementation and solution procedure is robust and shows the asymptotically quadratic rate of convergence associated with the Newton--Raphson scheme.
For the first time, specific constitutive laws for CPD have been presented.
Their numerical implementations has been discussed in detail together with analytical forms for their associated tangents.
The utility and reliability of the proposed strategy is illustrated via a broad range of numerical examples including three-dimensional problems at large deformations.

One logical extension of this work is to account for elasto-plasticity.
Another interesting extension of CPD, and our next immediate plan, is to consider thermomechanical problems.

\section*{Acknowledgement}

AJ and SF gratefully acknowledge the support provided by Scientific and Technological Research Council of Turkey (T{\"U}BITAK) Career Development Program, grant number 218M700.
PS and AM gratefully acknowledge the support provided by the EPSRC Strategic Support Package: Engineering of Active Materials by Multiscale/Multiphysics Computational Mechanics - EP/R008531/1.

\bibliography{library.bib}

\end{document}